\documentclass[twocolumn,showpacs,preprintnumbers,superscriptaddress,amsmath,amssymb]{revtex4}
\usepackage{graphicx}
\usepackage{dcolumn}
\usepackage{bm}
\usepackage{amssymb}
\usepackage{amssymb}
\def\>{\rangle}

\begin{document}

\title{Cooling and squeezing the fluctuations of a nanomechanical beam by indirect quantum feedback control}

\author{Jing Zhang}\email{jing-zhang@mail.tsinghua.edu.cn}
\affiliation{Advanced Science Institute, The Institute of Physical
and Chemical Research (RIKEN), Wako-shi, Saitama 351-0198, Japan}
\affiliation{Department of Automation, Tsinghua University,
Beijing 100084, P. R. China}
\author{Yu-xi Liu}
\affiliation{Advanced Science Institute, The Institute of Physical
and Chemical Research (RIKEN), Wako-shi, Saitama 351-0198,
Japan}\affiliation{CREST, Japan Science and Technology Agency
(JST), Kawaguchi, Saitama 332-0012, Japan}
\author{Franco Nori}
\affiliation{Advanced Science Institute, The Institute of Physical
and Chemical Research (RIKEN), Wako-shi, Saitama 351-0198,
Japan}\affiliation{CREST, Japan Science and Technology Agency
(JST), Kawaguchi, Saitama 332-0012, Japan} \affiliation{Center for
Theoretical Physics, Physics Department, Center for the Study of
Complex Systems, The University of Michigan, Ann Arbor,
Michigan 48109-1040, USA}%

\date{\today}

\begin{abstract}
We study cooling and squeezing the fluctuations of a
nanomechanical beam using quantum feedback control. In our model,
the nanomechanical beam is coupled to a transmission line
resonator via a superconducting quantum interference device
(SQUID). The leakage of the electromagnetic field from the
transmission line resonator is measured using homodyne detection.
This measured signal is then used to design a
quantum-feedback-control signal to drive the electromagnetic field
in the transmission line resonator. Although the control is
imposed on the transmission line resonator, this
quantum-feedback-control signal indirectly affects the thermal
motion of the nanomechanical beam via the inductive beam-resonator
coupling, making it possible to cool and squeeze the fluctuations
of the beam, allowing it to approach the standard quantum limit.

\end{abstract}


\maketitle

\section{Introduction}\label{s1}

Nanomechanical oscillators have recently attracted considerable
attention for their possible applications in quantum information
and quantum measurement (see, e.g.,
Refs.~\cite{Schwab,Blencowe,Ekinci,Cleland,Buks1,Sergey1,Mancini,Mahboob,Wei,Marshall,Huang,Gaidarzhy,Metzger,Naik,Gigan,Arcizet,Schliesser,Kippenberg,Corbitt,Bhattacharya,Kleckner,Cohadon,Poggio,Schliesser2,Rae,Marquardt,Genes,YDWang,MGrajcar,LTian,Naik2,Hopkins,Rae2}).
A nanomechanical oscillator is also a promising device for
studying macroscopic quantum effects in mechanical systems (see,
e.g.,
Refs.~\cite{Schwab,Blencowe,Ekinci,Cleland,Buks1,Sergey1,Mancini,Mahboob,Wei,Marshall}).
Using current experimental techniques~(see, e.g.,
Refs.~\cite{Ekinci,Huang,Gaidarzhy}), high-frequency
nanomechanical oscillators ($\omega/2\pi\sim1$ GHz) with quality
factors $Q$ in the range of $10^3$--$10^5$ can be realized at low
temperatures $T$ on the order of mK. When the vibrational energy
$\hbar\omega$ of the nanomechanical oscillator becomes smaller
than the thermal energy $k_B T$, the oscillator can be said to
work in the quantum regime.

To observe quantum behavior in nanomechanical oscillators, e.g.,
quantum fluctuations or squeezing effects, the oscillator must be
cooled to extremely low temperatures to approach the standard
quantum limit. There have been numerous studies, both theoretical
and experimental (see, e.g.,
Refs.~\cite{Metzger,Naik,Gigan,Arcizet,Schliesser,Kippenberg,Corbitt,Bhattacharya,Kleckner,Cohadon,Poggio,Schliesser2,Rae,Marquardt,Genes,YDWang,MGrajcar,LTian,Naik2,Hopkins,Rae2,Ouyang,Martin,PZhang,You,Hauss,Vinante,Xue,Teufel}),
investigating the cooling of the fluctuations of nanomechanical
oscillators. Many of these studies focus on optomechanical systems
(see, e.g.,
Refs.~\cite{Metzger,Naik,Gigan,Arcizet,Schliesser,Kippenberg,Corbitt,Bhattacharya,Kleckner,Cohadon,Poggio}),
where an oscillating cantilever or an oscillating micro-mirror is
modelled as a harmonic oscillator. There are two approaches in
optomechanical cooling: passive
cooling~\cite{Metzger,Naik,Gigan,Arcizet,Schliesser,Kippenberg,Corbitt,Bhattacharya}
and active cooling~\cite{Kleckner,Cohadon,Poggio}. In passive
cooling techniques, the mechanical oscillator is self-cooled by
the dynamical back-action, e.g., the radiation-pressure-induced
back-action~\cite{Naik,Gigan,Arcizet,Schliesser,Kippenberg,Corbitt,Bhattacharya}
coming from the mirror surface of the optical cavity. In fact, for
a high-finesse cavity, the photons reflected from the mirror of
the cavity transfer momentum and induce additional damping to the
mechanical oscillator. In active cooling techniques, the reflected
signal coming from the mechanical oscillator is sent to an
electronic circuit, e.g., a derivative circuit, to provide a
modulating signal, which is then used to control the back-action
force imposed on the mechanical oscillator. Since the cooling
effect can be actively controlled by tuning the feedback gain
obtained in the control circuit, this is called an active cooling
strategy.

Although it has recently been reported that ground-state
cooling~\cite{Schliesser2,Rae,Marquardt,Genes} could be realized
in optomechanical systems, it is difficult to observe the
macroscopic quantum effects of the mechanical oscillators in these
optomechanical systems using current experimental conditions. The
main difficulty comes from the fact that the characteristic
oscillating frequency of the mechanical oscillator in these
systems is not high enough (typically on the order of kHz or MHz),
and the corresponding effective temperature to observe the quantum
effects is extremely low (typically on the order of nK or $\mu$K),
which is difficult to realize in present-day experiments.

Besides optomechanical cooling, a nanomechanical oscillator can
also be embedded in an electronic circuit and cooled by coupling
it to an electronic system~\cite{YDWang,MGrajcar,LTian}. Possible
strategies include nanomechanical oscillators coupled to
superconducting single-electron transistors~\cite{Naik2,Hopkins},
quantum dots~\cite{Rae2,Ouyang}, Josephson-junction
superconducting circuits~\cite{Martin,PZhang,You,Hauss,Vinante},
or transmission line resonators~\cite{Xue,Teufel}. Compared with
mechanical oscillators in optical systems, a high-frequency
oscillator can be realized more easily in electronic systems.
Indeed, it has been reported that nanomechanical
beams~\cite{Huang,Gaidarzhy} with frequencies in the regime of GHz
have been realized, and these beams seem to be suitable for
integration in an electronic circuit. Since the effective
temperature of such a mechanical oscillator can be in the mK
regime, it should be possible to observe quantum behavior in this
case.

Like optomechanical systems, active feedback controls can be
introduced to cool the motions of the nanomechanical oscillators
in electronic systems. In the theoretical proposal in
Ref.~\cite{Hopkins}, a nanomechanical resonator is capacitively
coupled to a single-electron transistor to measure the position of
the resonator. The information obtained by the quantum measurement
is fed into a feedback circuit to obtain an output control signal,
which is then imposed on a feedback electrode to control the
motion of the resonator. In this strategy, the quantum measurement
and the designed feedback control introduce additional damping
effects on the resonator, which are helpful for cooling the motion
of the resonator.

In a recent experiment~\cite{Etaki}, a nanomechanical beam acted
as one side of a superconducting quantum interference device
(SQUID), and the voltage across the SQUID was measured which can
be used to detect the motion of the beam. Motivated by this
experiment, here we study the coupling between such a
system~\cite{Etaki} and a transmission line resonator. A
single-mode quantized electromagnetic field provided by this
resonator~\cite{You2} could be detected by a homodyne
measurement~\cite{Blais}. The quantization of this coupled
beam--SQUID--resonator system has been addressed in the literature
(see, e.g., Ref.~\cite{Buks2,Buks3,Buks4}), and theoretical
analysis shows that such a device can be used to detect the motion
of the beam~\cite{Buks4}. Here, we would like to concentrate on a
different problem: how to design a quantum feedback control from
the output signal of the homodyne detection to drive the motion of
the beam? Different from previous work, such as the one in
Ref.~\cite{Hopkins}, the quantum feedback control proposed here is
{\it imposed on the transmission line resonator, not on the beam,}
and indirectly controls the motion of the nanomechanical beam via
the coupling between the transmission line resonator and the beam.
By adiabatically eliminating the degrees of freedom of the SQUID
and the transmission line resonator, the designed feedback control
could introduce anharmonic terms in the effective Hamiltonian and
additional damping terms for the nanomechanical beam, {\it leading
to both cooling and squeezing} of the fluctuations of the beam.

This paper is organized as follows: in Sec.~\ref{s2}, we present a
coupled system composed of: (i) a transmission line resonator,
(ii) an rf-SQUID, and (iii) a nanomechanical beam. The open
quantum system model and the quantum weak measurement approach
used to design the feedback control are investigated in
Sec.~\ref{s3}. The quantum feedback control design is presented in
Sec.~\ref{s4}, and our main results about squeezing and cooling
the fluctuations of the beam are discussed in Sec.~\ref{s5}. The
conclusion and discussion of possible future work are presented in
Sec.~\ref{s6}.

\section{System model and Hamiltonian}\label{s2}

We mainly focus on a physical model in which a doubly-clamped
nanomechanical beam, a rf-SQUID, and a transmission line resonator
are inductively coupled (see, e.g.,
Ref.~\cite{Buks2,Buks3,Buks4}). The quantum electromechanical
circuit and the corresponding equivalent schematic diagram are
shown in Fig.~\ref{Fig of the coupled transmission
line-SQUID-nanomechanical beam system}. In this circuit, the
mechanical oscillator, i.e., the clamped nanomechanical beam, is
integrated into the rf-SQUID with a Josephson junction having a
critical current $I_{c}$ and a capacitance $C$. Here, the
displacement of the beam in the plane of the loop with a small
amplitude $x$ around its equilibrium position changes the area of
the loop, and thus influences the total magnetic flux $\Phi$
threading the loop. There is an applied external flux $\Phi_{e}$
threading the loop. The rf-SQUID interacts with a nearby
transmission line resonator (TLR), via their mutual inductance.
The additional magnetic flux provided by the quantized current in
the transmission line resonator is
\begin{eqnarray*}
\Phi_{\rm add}=\Phi_{T}(-i a+ia^{\dagger}),
\end{eqnarray*}
where $\Phi_{T}$ is a constant; $a$ and $a^{\dagger}$ are the
annihilation and creation operators of the quantized
electromagnetic field in the transmission line resonator. Here, we
ignore the small change of $\Phi_{T}$ caused by the oscillation of
the beam. For this superconducting circuit, the total magnetic
flux $\Phi$ threading the loop of the rf-SQUID is given by:
\begin{eqnarray*}
\Phi=\Phi_{e}+Blx+\Phi_{T}(-ia+ia^{\dagger})+LI,
\end{eqnarray*}
where $L$ and $I$ are, respectively, the self-inductance and the
current in the loop; $l$ is the effective length of the beam; and
$B$ is the magnetic field threading the loop of the rf-SQUID at
the location of the nanomechanical beam, and is assumed to be
constant in the region where the beam oscillates.

\begin{figure}[h]
\centerline{
\includegraphics[width=3.2in,height=2.5in, clip]{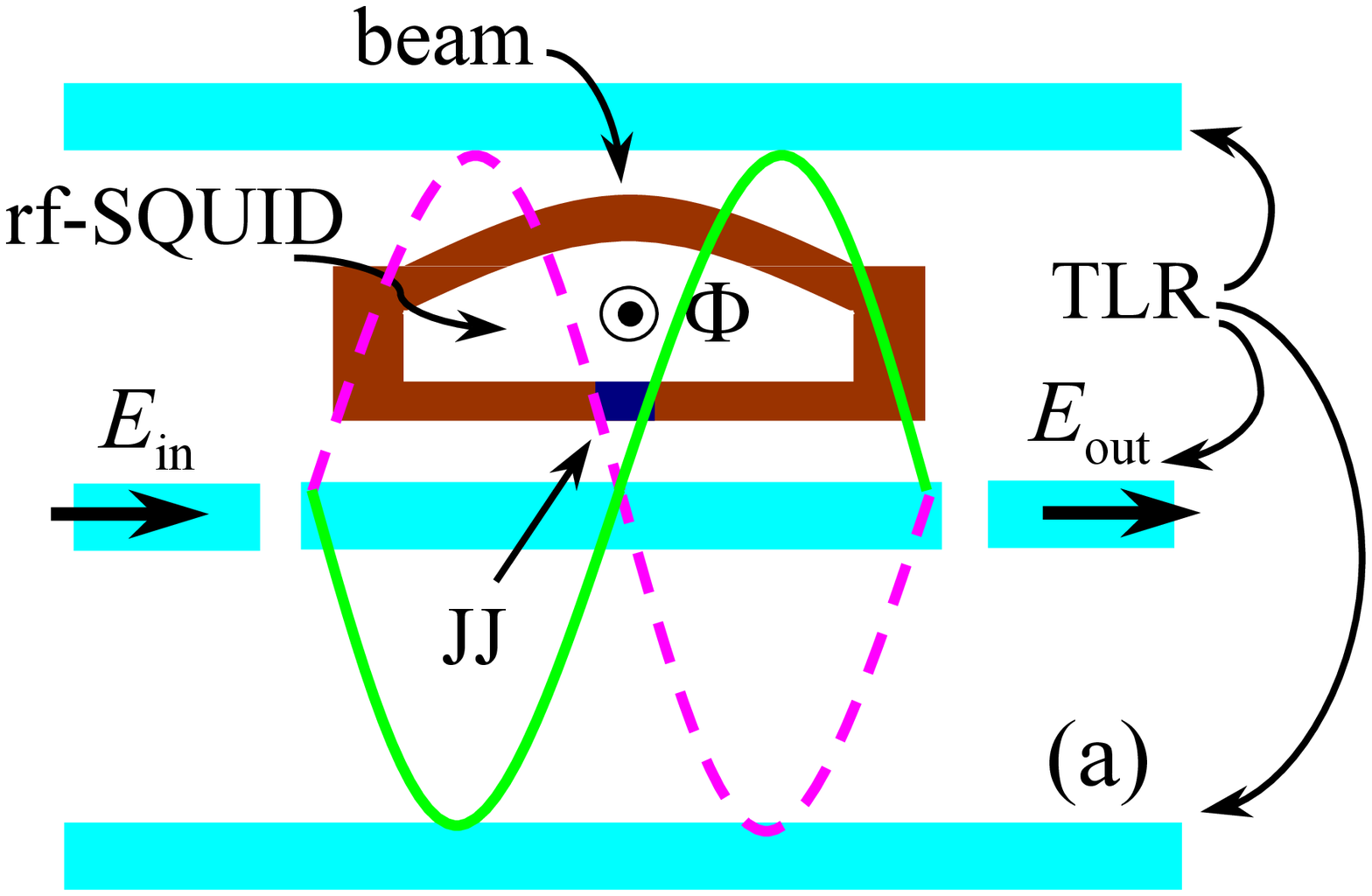}
} \centerline{
\includegraphics[width=2.8in,height=2.5in, clip]{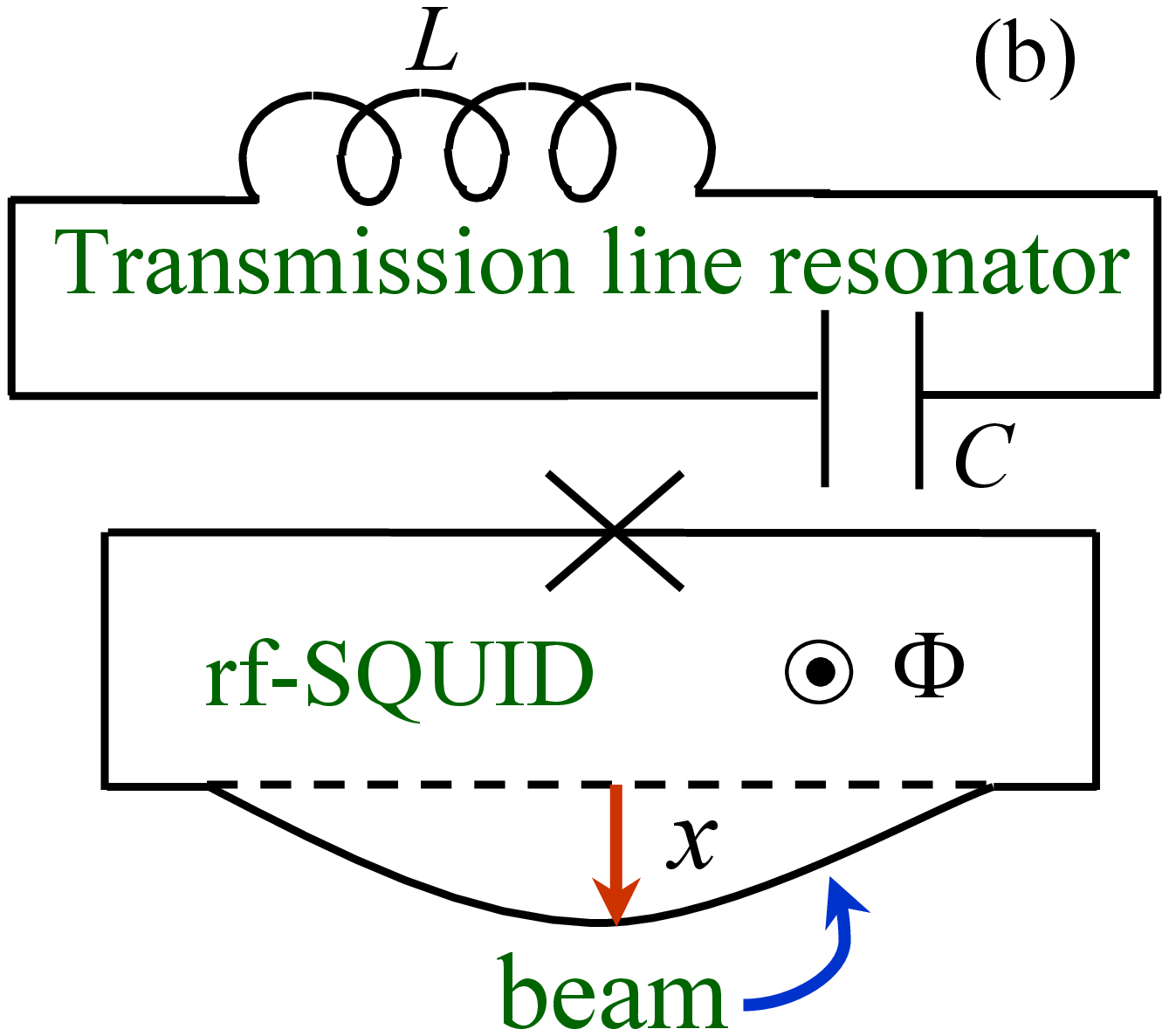}
} \caption{(color online) (a) Schematic diagram of a transmission
line resonator (TLR), in blue, and a SQUID-nanomechanical beam
system, in beige and dark blue. Here, ``JJ" represents a Josephson
junction, and $E_{\rm in}$, $E_{\rm out}$ are, respectively, the
input and output electromagnetic fields of the transmission line
resonator. (b) Diagram of the quantum circuit.}\label{Fig of the
coupled transmission line-SQUID-nanomechanical beam system}
\end{figure}

The total Hamiltonian of this coupled electromechanical system can
be written as~\cite{Buks3}:
\begin{equation}\label{Hamiltonian of the coupled resonator-SQUID-beam system}
H=H_0+H_1,
\end{equation}
with the Hamiltonians:
\begin{eqnarray}\label{H_0 and H_1}
H_0&=&\frac{p^2}{2m}+\frac{m\omega_{M}^2 x^2}{2}+\hbar\,\omega_{T}a^{\dagger}a+\hbar\,u(t)(a^{\dagger}+a),\label{H_0}\\
H_1&=&U_0\left[\phi-\phi_{e}-\kappa(-ia+ia^{\dagger})-\frac{2\pi B l}{\Phi_0}x\right]^2\nonumber\\
&&+2U_0\beta_{L}\cos\phi+\frac{q^2}{2C},\label{H_1}
\end{eqnarray}
where $\Phi_0$ is the flux quantum; $\omega_{M}$ is the
oscillating frequency of the doubly-clamped beam; $\phi$ and
$\phi_{e}$ are related to the normalized total flux and external
flux:
\begin{eqnarray*}
\phi=2\pi\left(\frac{\Phi}{\Phi_0}-\frac{1}{2}\right),\quad\phi_{e}=2\pi\left(\frac{\Phi_{e}}{\Phi_0}-\frac{1}{2}\right).
\end{eqnarray*}
The normalized system parameters $U_0$, $\beta_{L}$ and $\kappa$
in Eq.~(\ref{Hamiltonian of the coupled resonator-SQUID-beam
system}) are given by:
\begin{eqnarray*}
U_0&=&\frac{\Phi_0^2}{8\pi L},\quad\beta_{L}=\frac{2\pi L
I_{c}}{\Phi_0},\quad\kappa=\frac{2\pi\Phi_{T}}{\Phi_0}.
\end{eqnarray*}
The observables $p$ and $q$ in Eqs.~(\ref{H_0}) and (\ref{H_1})
are, respectively, the conjugate observables of $x$ and $\Phi$
representing the momentum of the beam and the charge on the
Josephson junction. The term $\hbar u(t)(a^{\dagger}+a)$ in
Eq.~(\ref{H_0}) is an interaction Hamiltonian between the
transmission line resonator and the external control field, where
the time-dependent function $u(t)$ can be designed according to
the desired goal.

When $\beta_{L}>1$ and
\begin{equation}\label{Condition for two-level }
\left|\phi_{e}+\frac{2\pi B l
x}{\Phi_0}+\kappa(-ia+ia^{\dagger})\right|=|\phi_{e}'|\ll 1,
\end{equation}
the Hamiltonian $H_1$ represents a double-well potential near
$\phi=0$, and the two lowest eigenstates, $|L\rangle$ and
$|R\rangle$, correspond to two current states with opposite
circulating currents in the loop of the rf-SQUID, which are far
separated from higher-energy eigenstates. At sufficiently low
temperatures, only the two lowest eigenstates $|L\rangle$,
$|R\rangle$ contribute. Thus, the rf-SQUID can be modelled as a
two-level system, and the Hamiltonian $H_1$ of the rf-SQUID can be
re-expressed as~\cite{Buks3}:
\begin{eqnarray*}
H_1=\frac{\hbar\,\epsilon}{2}\left(\phi_{e}+\frac{2\pi B
l}{\Phi_0}x+\kappa(-ia+ia^{\dagger})\right)\tilde{\sigma}_z-\frac{\hbar\,\Delta}{2}\tilde{\sigma}_x,
\end{eqnarray*}
where $\tilde{\sigma}_x$ and $\tilde{\sigma}_z$ are the $x$-axis
and $z$-axis Pauli operators in the basis of $|L\rangle$ and
$|R\rangle$; $\epsilon$, $\Delta$ are real parameters that
determine the energy difference between the two minima of the
double-well potential and the tunnelling amplitude between the
wells, respectively. Under the condition that
\begin{eqnarray*}
0<\left(\beta_{L}-1\right)\ll1,
\end{eqnarray*}
then $\epsilon$ and $\Delta$ can be approximately given by
\cite{RMigliore}:
\begin{equation}\label{Approximation of epsilon and Delta}
\epsilon=\frac{I_{c}\Phi_0}{\hbar\pi}\sqrt{6(\beta_L-1)},\quad\Delta=3U_0\left(1-\frac{1}{\beta_{L}}\right)^2.
\end{equation}

Letting the external flux $\phi_{e}=0$, we can rewrite the
Hamiltonian $H_1$ in the qubit basis as:
\begin{equation}\label{H_1 under the two-level approximation}
H_1=\frac{\hbar\,\omega_{S}}{2}\sigma_z+\frac{\hbar\,\pi\epsilon
Bl}{\Phi_0}x\sigma_x+\frac{\hbar\,\kappa\epsilon}{2}(-ia+ia^{\dagger})\sigma_x,
\end{equation}
where $\omega_{S}=\Delta$, and $\sigma_x$, $\sigma_z$ are the
corresponding $x$-axis and $z$-axis Pauli operators in the qubit
basis:
\begin{eqnarray*}
|+\rangle&=&\frac{\sqrt{2}}{2}|L\rangle+\frac{\sqrt{2}}{2}|R\rangle,\\
|-\rangle&=&\frac{\sqrt{2}}{2}|L\rangle-\frac{\sqrt{2}}{2}|R\rangle.
\end{eqnarray*}
Here, we assume that the oscillation frequency of the beam is high
enough such that $\omega_{M}$, $\omega_{S}$, and $\omega_{T}$ are
comparable. Then, under the rotating-wave approximation, and with
Eq.~(\ref{H_1 under the two-level approximation}), the total
Hamiltonian in Eq.~(\ref{Hamiltonian of the coupled
resonator-SQUID-beam system}) becomes:
\begin{eqnarray}\label{Hamiltonian of the coupled resonator-SQUID-beam system under the rotating-wave approximation}
H&=&\frac{\hbar\,\omega_{S}}{2}\sigma_z+\hbar\,\omega_{M}b^{\dagger}b+\hbar\,\omega_{T}a^{\dagger}a\nonumber\\
&&+\hbar\,u(t)(a^{\dagger}+a)+\hbar\,g_{MS}(b\sigma_++\sigma_-b^{\dagger})\nonumber\\
&&+\hbar\,g_{ST}(-ia\sigma_++i\sigma_-a^{\dagger}),
\end{eqnarray}
where the coupling strength $g_{MS}$ between the mechanical
oscillator and the rf-SQUID is:
\begin{eqnarray*}
g_{MS}=g_{\rm Mech-SQUID}=\frac{\pi \epsilon B
l}{\Phi_0\sqrt{2\hbar\,m\omega_{M}}},
\end{eqnarray*}
and the coupling strength $g_{ST}$ between the rf-SQUID and the
transmission line resonator is given by:
\begin{eqnarray*}
g_{ST}=g_{\rm SQUID-TLR}=\frac{\kappa\epsilon}{2\hbar}.
\end{eqnarray*}
The annihilation and creation operators $b$ and $b^{\dagger}$ of
the fundamental oscillating mode of the nanomechanical beam are
defined by:
\begin{eqnarray*}
b&=&\sqrt{\frac{m\omega_{M}}{2\hbar}}x+i\frac{1}{\sqrt{2\hbar\,m\omega_{M}}}p,\\
b^{\dagger}&=&\sqrt{\frac{m\omega_{M}}{2\hbar}}x-i\frac{1}{\sqrt{2\hbar\,m\omega_{M}}}p.
\end{eqnarray*}

Furthermore, let us assume that the frequencies of the rf-SQUID,
the beam, and the transmission line resonator satisfy the
conditions:
\begin{equation}\label{Large detuning assumption}
g_{MS}\ll\Delta_{MS}=\omega_{S}-\omega_{M},\quad
g_{ST}\ll\Delta_{ST}=\omega_{S}-\omega_{T}.
\end{equation}
Then, in this large-detuning regime~\cite{CPSun}, the following
transformation can be introduced to diagonalize the Hamiltonian
$H$ in Eq.~(\ref{Hamiltonian of the coupled resonator-SQUID-beam
system under the rotating-wave approximation}):
\begin{eqnarray*}
U=\exp\left[\frac{g_{MS}}{\Delta_{MS}}(b\sigma_+-b^{\dagger}\sigma_-)-\frac{g_{ST}}{\Delta_{ST}}(ia\sigma_++ia^{\dagger}\sigma_-)\right].
\end{eqnarray*}
In fact, under the condition given in Eq.~(\ref{Large detuning
assumption}), we can obtain an effective Hamiltonian:
\begin{eqnarray*}
H_{\rm eff}&=&UHU^{\dagger}\\
&\approx&\hbar\,\omega_{M} b^{\dagger} b+\hbar\,\omega_{T} a^{\dagger} a+\hbar\,u(t)(a^{\dagger}+a)\\
&&+\frac{\hbar\omega_{S}}{2}\sigma_z+\hbar\left(\frac{g_{MS}^2}{\Delta_{MS}}b^{\dagger}b+\frac{g_{ST}^2}{\Delta_{ST}}a^{\dagger}a\right)\sigma_z\\
&&+\hbar
\left(\frac{g_{MS}g_{ST}}{\Delta_{MS}}+\frac{g_{MS}g_{ST}}{\Delta_{ST}}\right)(-ib
a^{\dagger}+ib^{\dagger}a)\sigma_z,
\end{eqnarray*}
by expanding $UHU^{\dagger}$ to first order in
$g_{MS}/\Delta_{MS}$ and $g_{ST}/\Delta_{ST}$.

\section{Interaction between the system and its environment}\label{s3}

A real physical system inevitably interacts with the external
degrees of freedom in the environment. Such interactions introduce
noise to the system. There are three kinds of noise that should be
considered here: the thermal noises on the nanomechanical beam and
the transmission line resonator, as well as the electromagnetic
fluctuations on the rf-SQUID caused by the nearby electromagnetic
elements.

The interaction Hamiltonians between the transmission line
resonator, the beam, the rf-SQUID and their environments can be
described as:
\begin{eqnarray*}
H_{\rm int}^{(1)}&=&(a+a^{\dagger})X_{\rm bath}^{(1)},\\
H_{\rm int}^{(2)}&=&(b+b^{\dagger})X_{\rm bath}^{(2)},\\
H_{\rm int}^{(3)}&=&(M_{\phi}\sigma_z+M_{r}\sigma_x)X_{\rm
bath}^{(3)},
\end{eqnarray*}
where $X_{\rm bath}^{(1)}$, $X_{\rm bath}^{(2)}$ and $X_{\rm
bath}^{(3)}$ are, respectively, the environmental operators
interacting with the transmission-line resonator, the beam and the
rf-SQUID; $M_{\phi}$ and $M_{r}$ are two constants which determine
the dephasing and relaxation rates of the rf-SQUID. Furthermore,
let us consider a bosonic model of the environment, and assume
that the interactions between the system degrees of freedom and
the environmental degrees of freedom are linear interactions.
Then, under the rotating-wave approximation and the Markovian
approximation, we can obtain the following quantum stochastic
differential equation for a system observable $X$ (see
Appendix~\ref{Derivation of the quantum differential equation} for
the derivation):
\begin{eqnarray}\label{Quantum stochastic differential equation of the coupled resonator-SQUID-beam system}
dX&=&-\frac{i}{\hbar}[X,H_{\rm eff}]dt+\frac{\gamma_{S}}{2}M_{\phi}^2[\sigma_z,[X,\sigma_z]]\nonumber\\
&&+\frac{\gamma_{S}}{2}M_{r}^2\left(\sigma_+[X,\sigma_-]+[\sigma_+,X]\sigma_-\right)\nonumber\\
&&+\frac{\gamma_{M}}{2}(\bar{n}_{M}+1)\left(b^{\dagger}[X,b]+[b^{\dagger},X]b\right)\nonumber\\
&&+\frac{\gamma_{M}}{2}\bar{n}_{M}\left(b[X,b^\dagger]+[b,X]b^{\dagger}\right)\nonumber\\
&&+\frac{\gamma_{T}}{2}\left(a^{\dagger}[X,a]+[a^{\dagger},X]a\right)\nonumber\\
&&+\sqrt{\eta\gamma_{T}}dA_{\rm
in}^{\dagger}[X,a]+\sqrt{\eta\gamma_{T}}[a^{\dagger},X]dA_{\rm
in},
\end{eqnarray}
where $\gamma_{M}$, $\gamma_{S}$, $\gamma_{T}$ are the damping
rates of the mechanical beam, the rf-SQUID and the transmission
line resonator under the Markovian approximation;
\begin{equation}\label{nbarm}
\bar{n}_{M}=\frac{1}{e^{\hbar\omega_{M}/k_{B}T}-1}
\end{equation}
is the average photon number of the beam in thermal equilibrium
with the environment at temperature $T$. To simplify our
discussions, when Eq.~(\ref{Quantum stochastic differential
equation of the coupled resonator-SQUID-beam system}) was derived,
we neglected environment-induced thermal excitations on the
transmission line resonator and the rf-SQUID (these excitations
could indeed be neglected with the parameters given in
Eqs.~(\ref{System parameters1}) and (\ref{System parameters2}) in
Sec.~\ref{s5}). The leakage of the transmission line resonator
could be detected using a homodyne detection with detection
efficiency $\eta$, where $dA_{\rm in}$ represents a quantum Wiener
noise~\cite{Gardiner} satisfying:
\begin{eqnarray*}
&dA_{\rm in}^{\dagger}dA_{\rm in}=dA_{\rm in}dA_{\rm in}=dA_{\rm in}^{\dagger}dA_{\rm in}^{\dagger}=0,&\\
&dA_{\rm in}dA_{\rm in}^{\dagger}=dt.&
\end{eqnarray*}
Here, we only keep the fluctuation terms caused by the measurement
and average over the other fluctuations, because the evolution of
the coupled beam-SQUID-resonator system is conditioned on the
measurement output, which depends on the measurement-induced
fluctuations. The corresponding measurement output of the homodyne
detection can be expressed as~\cite{Buks4}:
\begin{equation}\label{Measurement output of the homodyne detection}
dY_{t}=\sqrt{\eta\gamma_{T}}(a^{\dagger}+a)+\left(dA_{\rm
in}+dA_{\rm in}^{\dagger}\right).
\end{equation}
Note that this measurement output depends on the input noise and
the electromagnetic field of the transmission line resonator.

\section{Quantum filtering and quantum feedback control}\label{s4}

There are two possible ways to design a quantum feedback control
protocol~\cite{Mabuchi,Jacobs1} based on the measurement output.
One approach is to directly feed back the output signal to design
the quantum feedback control signal, which leads to the Markovian
quantum feedback control~\cite{Wiseman1}. Another approach, which
is called quantum Bayesian feedback control~\cite{Doherty1}, can
be divided into two steps: the first step is to find a so-called
quantum filtering equation~\cite{Belavkin,Bouten,NYamamoto1} to
give an estimate of the state of the system from the measurement
output; the second step is to design a feedback control signal
based on the estimated state. The possibility for a ``control
problem" to be divided into these two steps, i.e., a separate
filtering step and a control step, is called the separation
principle in control theory, which has recently been developed for
quantum control systems~\cite{Bouten}. Compared with the Markovian
quantum feedback control, the quantum Bayesian feedback control
can be applied to more general systems. In our proposal, we will
design the control using Bayesian feedback control.

Based on quantum filtering theory, which has been well developed
in the literatures~\cite{Belavkin,Bouten}, we can obtain the
following stochastic master equation for the estimated state
$\tilde{\rho}$:
\begin{eqnarray}\label{Stochastic master equation for the beam and the transmission line resonator}
d\tilde{\rho}&=&-\frac{i}{\hbar}[H_{\rm
eff},\tilde{\rho}]dt+\gamma_M\bar{n}_M\mathcal{D}\left[b^{\dagger}\right]\tilde{\rho}dt\nonumber\\
&&+\gamma_M\left(\bar{n}_M+1\right)\mathcal{D}[b]\tilde{\rho}dt+\gamma_S
M_{\phi}^2
\mathcal{D}[\sigma_z]\tilde{\rho}dt\nonumber\\
&&+\gamma_S
M_r^2\mathcal{D}[\sigma_-]\tilde{\rho}dt+\gamma_T\mathcal{D}[a]\tilde{\rho}dt\nonumber\\
&&+\sqrt{\eta\gamma_T}\mathcal{H}[a]\tilde{\rho}\left(dY_t-\sqrt{\eta\gamma_T}\langle\left(a^{\dagger}+a\right)\rangle
dt\right),
\end{eqnarray}
where
\begin{equation}\label{Definition of the quantum filtering state}
\tilde{\rho}=E(\rho|\mathcal{Y}_{t})
\end{equation}
is defined as the conditional expectation of the density operator
$\rho$ for the coupled beam-SQUID-resonator system under the von
Neumann algebra~\cite{Bouten}:
\begin{equation}\label{Measurement out von Neumann algebra}
\mathcal{Y}_{t}={vN}\left\{Y_{s}|t_0\leq s\leq t\right\},
\end{equation}
spanned by the measurement outputs; $\langle A\rangle={\rm
tr}(A\tilde{\rho})$ is the average of $A$ under $\tilde{\rho}$;
and the superoperators $\mathcal{D}[c]\tilde{\rho}$ and
$\mathcal{H}[c]\tilde{\rho}$ are defined by:
\begin{eqnarray*}
\mathcal{D}[c]\tilde{\rho}&=&c\tilde{\rho}c^{\dagger}-\frac{1}{2}c^{\dagger}c\tilde{\rho}-\frac{1}{2}\tilde{\rho}c^{\dagger}c,\\
\mathcal{H}[c]\tilde{\rho}&=&c\tilde{\rho}+\tilde{\rho}c^{\dagger}-\langle
\left(c+c^{\dagger}\right)\rangle\tilde{\rho}.
\end{eqnarray*}
The increment
\begin{equation}\label{Wiener increment}
dW=dY_{t}-\sqrt{\eta\gamma_{T}}\langle\left(a^{\dagger}+a\right)\rangle
dt
\end{equation}
in Eq.~(\ref{Stochastic master equation for the beam and the
transmission line resonator}) is the innovation updated by the
quantum measurement, which has been proved to be a classical
Wiener increment satisfying:
\begin{eqnarray*}
E(dW)=0,\quad (dW)^2=dt.
\end{eqnarray*}
for homodyne detection (see, e.g., Ref.~\cite{Bouten}).

The von Neumann algebra $\mathcal{Y}_t$ defined in
Eq.~(\ref{Measurement out von Neumann algebra}) represents the
information obtained by the quantum measurement up to time $t$;
thus the conditional expectation $\tilde{\rho}$ defined in
Eq.~(\ref{Definition of the quantum filtering state}) is the best
estimate of the system's state obtained from the measurement
output.

Generally, the stochastic master equation, i.e., the filter
equation, is difficult to solve. However, for the system we
discuss here, the stochastic master equation (\ref{Stochastic
master equation for the beam and the transmission line resonator})
is equivalent to a set of closed equations under the semiclassical
approximation (see Eqs.~(\ref{Maxwell-Bloch-type equation for the
coupled beam-SQUID-resonator system}) and (\ref{Equations for the
variances and covariance of the transmission line resonator}) in
Appendix~\ref{Derivation of the reduced stochastic master equation
for the beam}). This set of equations can be integrated by a Data
Acquisition Processor (DAP) (see, e.g., Ref.~\cite{Mabuchi}),
which is composed of a Digital Signal Processor (DSP) and
analog/digital and digital/analog signal converters. Such a Data
Acquisition Processor works as an integral estimator of the
dynamics of the system state and gives the output signals $\langle
x_T\rangle$ and $\langle p_T\rangle$. The output signals of the
estimator are fed into a feedback controller (a linear
amplification element) to obtain the following feedback control
signal:
\begin{equation}\label{Linear feedback control}
u(t)=\sqrt{\frac{2}{\hbar}}\left(-\upsilon_{x}\langle
x_{T}\rangle+\upsilon_{p}\langle p_{T}\rangle\right),
\end{equation}
where $\upsilon_{x}$ and $\upsilon_{p}$ are the feedback control
gains that can be chosen according to the desired goal, and $x_T$,
$p_T$ are the normalized position and momentum operators of the
transmission line resonator defined by:
\begin{equation}\label{Normalized position and momentum of the transmission line resonator}
x_T=\sqrt{\frac{\hbar}{2}}\left(a^{\dagger}+a\right),\quad
p_T=\sqrt{\frac{\hbar}{2}}\left(-ia+ia^{\dagger}\right).
\end{equation}

We replace $u(t)$ in Eq.~(\ref{Quantum stochastic differential
equation of the coupled resonator-SQUID-beam system}) and
Eq.~(\ref{Stochastic master equation for the beam and the
transmission line resonator}) by Eq.~(\ref{Linear feedback
control}) to obtain new dynamical equations. In this case, we
indeed control, simultaneously, the evolutions of the transmission
line resonator and the estimator. The schematic diagram of the
feedback control circuit is shown in Fig.~\ref{Fig of the feedback
control circuit}. From the definition (\ref{Definition of the
quantum filtering state}) of $\tilde{\rho}$, it can be shown that
the control of the coupled system given by Eq.~(\ref{Quantum
stochastic differential equation of the coupled
resonator-SQUID-beam system}) is equivalent to the control of the
estimator given by Eq.~(\ref{Stochastic master equation for the
beam and the transmission line resonator}). Thus, in the following
discussion, we will focus on how to control the quantum filtering
equation (\ref{Stochastic master equation for the beam and the
transmission line resonator}).
\begin{figure}
\includegraphics[width=3.3in,height=3in, clip]{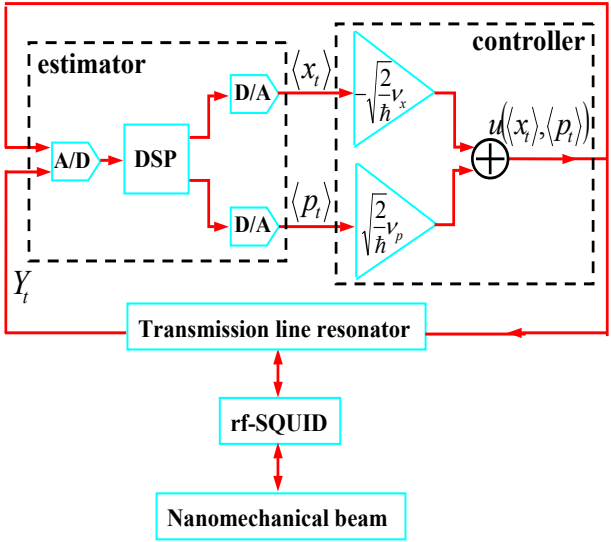}
\caption{ (color online) Schematic diagram of the feedback control
circuit. The ``DSP" denotes a Digital Signal Processor (DSP) which
works as the integral estimator of the system state by solving
Eqs.~(\ref{Maxwell-Bloch-type equation for the coupled
beam-SQUID-resonator system}) and (\ref{Equations for the
variances and covariance of the transmission line resonator}). The
``A/D" and ``D/A" represent the analog/digital and digital/analog
signal converters. The output of the estimator is fed into a
linear amplifier circuit to obtain a control signal which is
further used to drive the input electromagnetic field of the
transmission line resonator and the input of the
estimator.}\label{Fig of the feedback control circuit}
\end{figure}

If the damping rates $\gamma_S$ and $\gamma_{T}$ of the rf-SQUID
and the transmission line resonator are large enough such that
\begin{equation}\label{Adiabatic elimination condition}
\gamma_{S},\,\gamma_{T}\gg\gamma_{M}\bar{n}_{M},
\end{equation}
we can adiabatically eliminate~\cite{Walls,Steck} the degrees of
freedom of the rf-SQUID and the transmission line resonator to
obtain the following reduced stochastic master equation and the
measurement output for the nanomechanical beam (see
Appendix~\ref{Derivation of the reduced stochastic master equation
for the beam} for the derivation):
\begin{eqnarray}\label{Stochastic master equation of the beam}
d\tilde{\rho}_{M}&=&-\frac{i}{\hbar}[H^{(M)}_{\rm eff},\tilde{\rho}_{M}]dt+\gamma_{M}\bar{n}_{M}\mathcal{D}[b^{\dagger}]\tilde{\rho}_{M} dt\nonumber\\
&&+\gamma_{M}(\bar{n}_{M}+1)\mathcal{D}[b]\tilde{\rho}_{M} dt\nonumber\\
&&+\gamma_{T}\mathcal{D}[C_1 b +C_2 b^{\dagger}]\tilde{\rho}_{M} dt\nonumber\\
&&+\sqrt{\eta\gamma_{T}}\mathcal{H}[C_1 b+C_2 b^{\dagger}]\tilde{\rho}_{M} dW,\nonumber\\
dY_{t}&=&\sqrt{\eta\gamma_{T}}\langle\left(\alpha_{x}
b+\alpha_{x}^* b^{\dagger}\right)\rangle dt+dW,
\end{eqnarray}
where the reduced Hamiltonian $H^{(M)}_{\rm eff}$ is given by:
\begin{eqnarray}\label{Hmeff}
H^{(M)}_{\rm eff}&=&\hbar\,\omega_{M} b^{\dagger} b+\hbar\,\xi_{M} b^2 +\hbar\,\xi_{M}^* b^{\dagger\,2}\nonumber\\
&&+\hbar\,\tilde{u}(t)(\alpha_{x} b+\alpha_{x}^* b^{\dagger}),
\end{eqnarray}
and the reduced effective control on the beam is:
\begin{eqnarray*}
\tilde{u}(t)&=&-\upsilon_{x}\langle\left(\alpha_{x} b+\alpha_{x}^*
b^{\dagger}\right)\rangle+\upsilon_{p}\langle\left(\alpha_{p}
b+\alpha_{p}^* b^{\dagger}\right)\rangle.
\end{eqnarray*}
The parameters $\alpha_{x}$, $\alpha_{p}$, $\xi_{M}$ can be
expressed as:
\begin{eqnarray*}
\alpha_{x}&=&C_1+C_2^*,\\
\alpha_{p}&=&-iC_1+i C_2^*,\\
\xi_{M}&=&\omega_{T} C_2^* C_1-ig_{MT} C_2^*,
\end{eqnarray*}
where $g_{MT}$ is given by Eq.~(\ref{Reduced coupling strength
between the beam and the resonator}) and $C_1$, $C_2$ are given
by:
\begin{eqnarray}\label{C1C2xi}
C_1\left(v_{x},v_{p}\right)&=&\frac{g_{MT}}{\chi}\left[\left(\upsilon_{p}+\frac{\gamma_{T}}{2}\right)-i(-\upsilon_{x}+\omega_{T})\right],\nonumber\\
C_2\left(v_{x},v_{p}\right)&=&\frac{g_{MT}}{\chi}(\upsilon_{p}-i\upsilon_{x}),\nonumber\\
\chi\left(v_{x},v_{p}\right)&=&\frac{\gamma_T}{4}(\gamma_{T}+4\upsilon_{p})+\omega_{T}(\omega_{T}-2\upsilon_{x}).
\end{eqnarray}

As shown in Eq.~(\ref{Hmeff}), there is a two-photon term
$\hbar\xi_{M} b^2+\hbar\xi_{M}^*b^{\dagger\,2}$ in the effective
Hamiltonian $H_{\rm eff}^{(M)}$, which leads to squeezing in the
fluctuations of the beam. Without the quantum feedback control,
i.e., $\upsilon_{x}=\upsilon_{p}=0$, $\xi_{M}$ would be zero and
the two-photon term vanishes.

Equation (\ref{Stochastic master equation of the beam}) shows that
the quantum measurement and feedback control introduce extra
damping and fluctuation terms for the beam (the third and fourth
lines in Eq.~(\ref{Stochastic master equation of the beam})).
These damping terms are important for squeezing and cooling the
fluctuations of the beam.

\section{Squeezing and cooling the fluctuations of the nanomechanical beam}\label{s5}
In order to study the squeezing and cooling effects on the
nanomechanical beam induced by the quantum feedback control, let
us first define the normalized position and momentum operators of
the nanomechanical beam:
\begin{eqnarray*}
x_{M}=\sqrt{\frac{\hbar}{2}}(b+b^{\dagger}),\quad
p_{M}=\sqrt{\frac{\hbar}{2}}(-i b+ib^{\dagger}).
\end{eqnarray*}
Then, from the reduced stochastic master equation (\ref{Stochastic
master equation of the beam}), we can study the evolutions and the
corresponding stationary values of the variances
\begin{equation}\label{Variances of the normalized position and momentum of the beam}
V_{x_M}=\langle x_{M}^2\rangle-\langle x_{M}\rangle^2,\quad
V_{p_M}=\langle p_{M}^2\rangle-\langle p_{M}\rangle^2
\end{equation}
of $x_{M}$ and $p_{M}$.

\subsection{Squeezing}\label{s51}
The nanomechanical beam can be described by the conjugate
variables $x_M$ and $p_M$, and is in a squeezed state if the
corresponding variances of these variables defined in
Eq.~(\ref{Variances of the normalized position and momentum of the
beam}) satisfy $V_{x_M}<\hbar/2$ or $V_{p_M}<\hbar/2$, i.e., the
variance of one of the two conjugate variables is below the
standard quantum limit (see, e.g., Refs.~\cite{Hu,Hu2,Zagoskin}).
Owing to the uncertainty principle which requires that
$V_{x_M}V_{p_M}\geq\hbar^2/4$, squeezing the fluctuations of one
of the two conjugate variables would lead to the dispersion of the
fluctuations of the other conjugate variable.

If we choose the feedback control gains $\upsilon_{x}$ and
$\upsilon_{p}$ to satisfy the conditions
\begin{eqnarray}\label{Region of the control parameters}
0&<&\frac{\gamma_{T}}{\gamma_{T}+4\upsilon_{p}}\ll 1,\nonumber\\
0&<&\frac{\omega_{T}-2\upsilon_{x}}{\omega_{T}}\ll 1,\nonumber\\
0&<&\frac{\gamma_T g_{MT}^2\omega_{T}^2}{\omega_{M}}\ll \chi^2,
\end{eqnarray}
then the stationary variances $V^{c}_{x_M}$ and $V^{c}_{p_M}$ can
be estimated as:
\begin{equation}\label{Controlled stationary variances}
V_{x_M}^{c}\approx\frac{\hbar}{2}\sqrt{\frac{\xi}{\eta}},\quad
V_{p_M}^{c}\approx\frac{\hbar}{2}\frac{1}{\sqrt{\xi\eta}},
\end{equation}
where
\begin{equation}\label{xi and xim}
\xi=\frac{\omega_{M}-2{\rm Re}\,\xi_{M}}{\omega_{M}+2{\rm
Re}\,\xi_{M}},\quad {\rm Re}\,\xi_{M}\approx\frac{\omega_{T}
g^2_{MT}}{\chi^2}(\upsilon_{p}^2-\upsilon_{x}^2).
\end{equation}
It is shown in Eq.~(\ref{Controlled stationary variances}) that
the parameter $\xi$ determines the tradeoff of the squeezing
effects between $V_{x_M}$ and $V_{p_M}$. When $\xi>1$, the
fluctuation of the momentum $p_{M}$ of the nanomechanical beam is
squeezed. However, when $\xi<1$, the fluctuation of the position
$x_{M}$ of the nanomechanical beam is squeezed. The parameter
$\eta$, i.e., the measurement efficiency of the homodyne
detection, determines the minimum uncertainty that can be reached.
For a quantum weak measurement with a high efficiency $\eta$ such
that
\begin{eqnarray*}
\frac{1}{(2\bar{n}_{M}+1)^2}<\eta<1,
\end{eqnarray*}
where $\bar{n}_{M}$ is the thermal excitation number of the beam
given in Eq.~(\ref{nbarm}), it can be verified that
\begin{eqnarray*}
\frac{\hbar^2}{4}&<&V_{x_M}^{c} V_{p_M}^{c}\approx\frac{\hbar^2}{4\eta}\\
&<&V_{x_M}^{uc}V_{p_M}^{uc}=\hbar^2\left(\bar{n}_{M}+\frac{1}{2}\right)^2,
\end{eqnarray*}
where $V^{uc}_{x_M}$ and $V^{uc}_{p_M}$ are the ``uncontrolled"
stationary variances that are obtained from Eq.~(\ref{Stochastic
master equation of the beam}) by letting
$\upsilon_{x}=\upsilon_{p}=0$. It should be pointed out that {\it
the product of the uncertainties of $x_{M}$ and $p_{M}$ given by
Eq.~(\ref{Controlled stationary variances}), i.e.,
\begin{eqnarray*}
V_{x_M}^{c} V_{p_M}^{c}\approx\frac{\hbar^2}{4\eta},
\end{eqnarray*}
corresponds to the Heisenberg uncertainty limit of a quantum
system under imperfect quantum weak measurements} (see, e.g.,
Ref.~\cite{NYamamoto2}). When the measurement efficiency $\eta$
tends to unity, the traditional Heisenberg uncertainty limit
$\hbar^2/4$ is recovered.

To show the validity of our strategy, let us show some numerical
examples. The system parameters are chosen as~\cite{Buks3}:
\begin{eqnarray}\label{System parameters1}
&L=3.38\times10^{-11}\,\,{\rm H},\quad C=7.4\times 10^{17}\,\,{\rm F},&\nonumber\\
&I_{c}=10\,\,{\rm \mu A},\quad m=10^{-16}\,\,{\rm kg},\quad \eta=0.6,&\nonumber\\
&\omega_{M}/2\pi=1\,\,{\rm GHz},\quad Bl=1\,\,{\rm T}\times{\rm \mu m},&\nonumber\\
&Q=10^4,\quad T=100\,\,{\rm mK},\quad\phi_e=0,&\nonumber\\
&\gamma_{S}/2\pi=100\,\,{\rm MHz},\quad\gamma_{T}/2\pi=20\,\,{\rm MHz},&\nonumber\\
&g_{ST}/2\pi=20\,\,{\rm MHz},\quad\omega_{T}/2\pi=4.3\,\,{\rm
GHz}.&
\end{eqnarray}
From the above parameters, it can be calculated that
\begin{eqnarray}\label{System parameters2}
&\gamma_{M}/2\pi\approx 0.1\,\,{\rm MHz},\quad\omega_{S}/2\pi\approx 6.3\,\,{\rm GHz}&\nonumber\\
&g_{MS}/2\pi\approx73\,\,{\rm MHz},\quad g_{MT}/2\pi\approx
4.9\,\,{\rm MHz}.&
\end{eqnarray}
In our numerical results, the stationary variances are calculated
from the dynamic equation (\ref{Stochastic master equation of the
beam}), and the feedback control parameters $\upsilon_{x}$ and
$\upsilon_{p}$ are chosen to satisfy Eq.~(\ref{Region of the
control parameters}). In fact, in Fig.~\ref{Fig of the squeezing
of the normalized position operator}, we choose $\upsilon_{x}$ and
$\upsilon_{p}$ such that
\begin{equation}\label{Region of the control parameters for squeezing the position}
\upsilon_{x}=0.5\,\omega_{T},\quad
0.5\leq\frac{\upsilon_{p}}{\omega_{T}}\leq 1.
\end{equation}
In this case, we have $\upsilon_{p}^2\geq\upsilon_{x}^2$. Then,
from Eqs.~(\ref{Controlled stationary variances}) and (\ref{xi and
xim}), it can be verified that $V_{x_M}^{c}\leq V_{p_M}^c$, which
coincides with the numerical results in Fig.~\ref{Fig of the
squeezing of the normalized position operator}(a) (the blue dashed
line for $V_{x_M}^{c}/\hbar$ is below the green triangular line
for $V_{p_M}^c/\hbar$). It means that the fluctuation of the
position of the beam is squeezed. Meanwhile, the numerical results
in Fig.~\ref{Fig of the squeezing of the normalized position
operator}(b) show that the variances $V_{x_M}^{c}$ and
$V_{p_M}^{c}$ under control are much smaller than the variances
$V_{x_M}^{uc}$ and $V_{p_M}^{uc}$ without control, which means
that the designed feedback control reduces the variances of the
position and momentum of the beam. The product of the variances
under control $V_{x_M}^{c}V_{p_M}^{c}$ could even be reduced to be
close to the Heisenberg uncertainty limit $\hbar^2/4$.

\begin{figure}[h]
\centerline{
\includegraphics[width=3.2in,height=2.5in, clip]{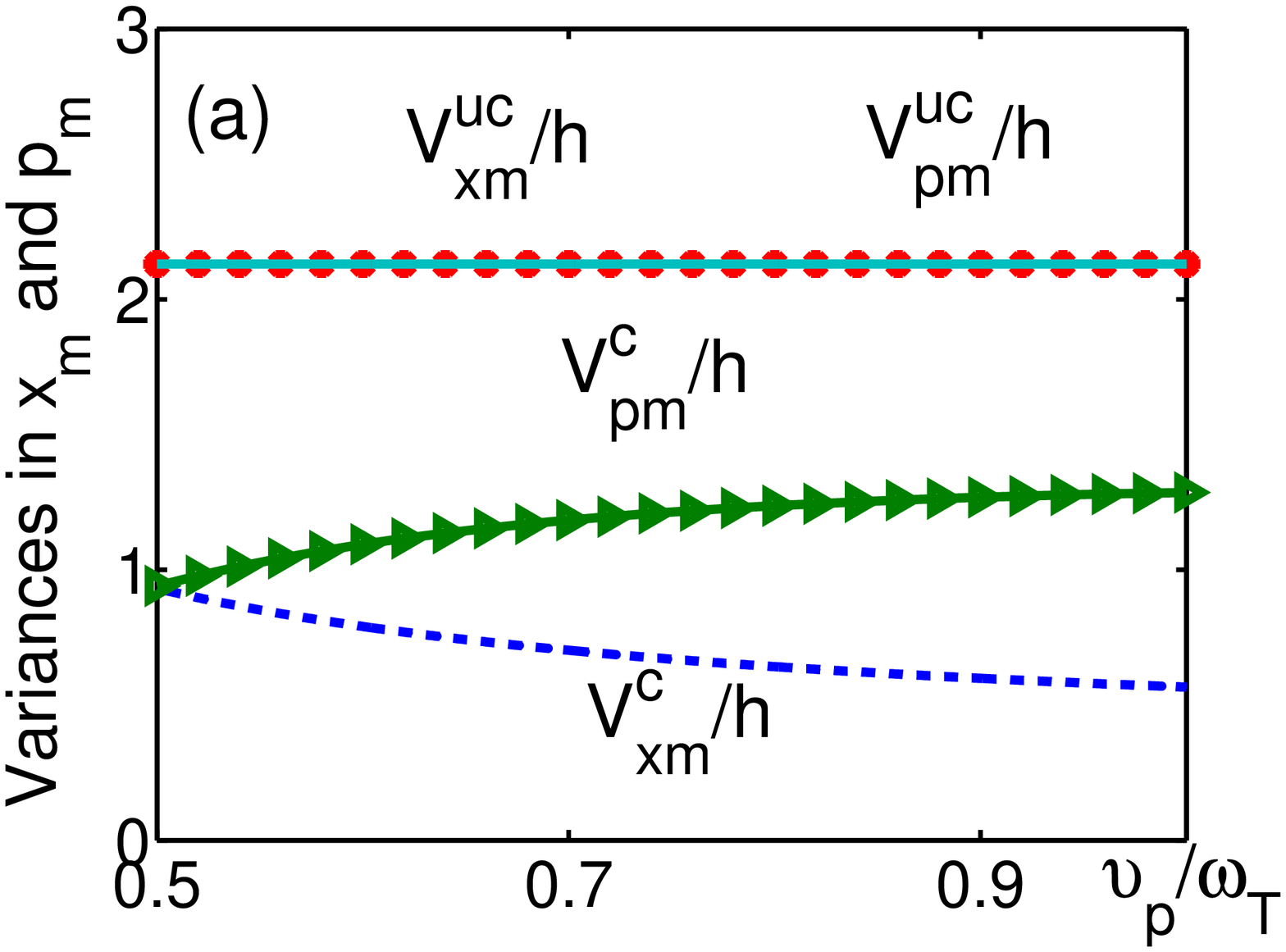}
} \centerline{
\includegraphics[width=3.2in,height=2.5in, clip]{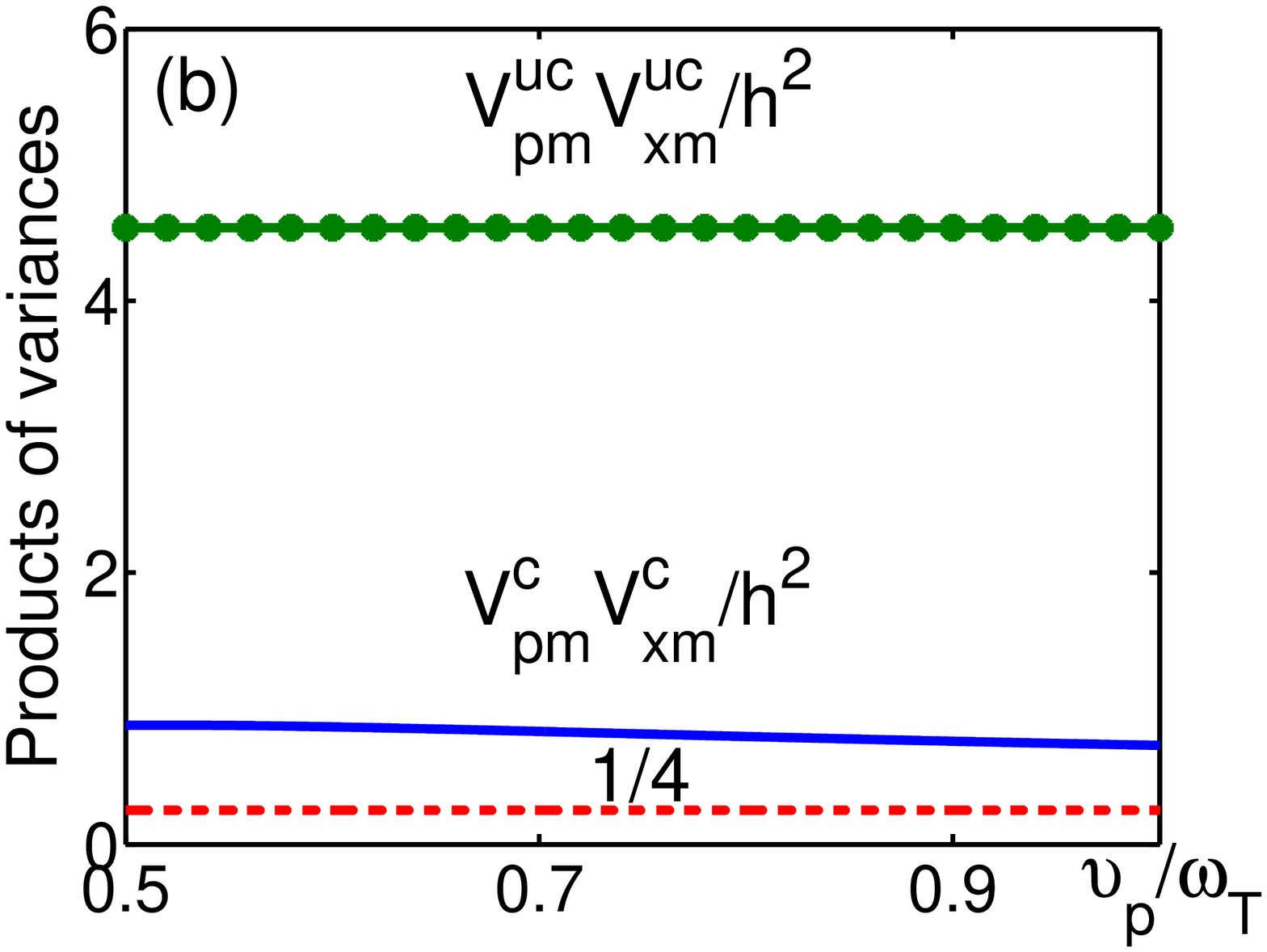}
} \caption{(color online) Squeezing the position-fluctuations of
the beam with the parameters given in Eqs.~(\ref{System
parameters1}), (\ref{System parameters2}), and (\ref{Region of the
control parameters for squeezing the position}). (a) Variances of
the position and momentum of the beam (in units of $\hbar$) versus
the normalized control parameter
$\tilde{\upsilon}_p=\upsilon_p/\omega_T$. Recall that $\upsilon_p$
is the linear feedback gain (see, Eq.~(\ref{Linear feedback
control})). The green line with triangles and the blue dashed line
represent, respectively, the controlled variances of the position
$V_{x_M}^c/\hbar$ and momentum $V_{p_M}^c/\hbar$ of the beam. The
red line with asterisks and the blue solid line represent the
uncontrolled variances $V_{x_M}^{uc}/\hbar$ and
$V_{p_M}^{uc}/\hbar$ of the beam, which coincide because the beam
without control is in a coherent thermal state with equal
variances of position and momentum. (b) Products of the variances
(in units of $\hbar^2$) versus the normalized control parameter
$\tilde{\upsilon}_p=\upsilon_p/\omega_T$. The green line with
asterisks and the blue solid line denote the uncontrolled
trajectory $V_{x_M}^{uc}V_{p_M}^{uc}/\hbar^2$ and the controlled
trajectory $V_{x_M}^c V_{p_M}^c/\hbar^2$, respectively. The red
dashed line at $1/4$ represents the Heisenberg uncertainty limit
$\hbar^2/4$.}\label{Fig of the squeezing of the normalized
position operator}
\end{figure}

In Fig.~\ref{Fig of the squeezing of the normalized momentum
operator}, we choose $\upsilon_{x}$ and $\upsilon_{p}$ such that
\begin{equation}\label{Region of the control parameters for squeezing the momentum}
\upsilon_{x}=0.5\,\omega_{T},\quad
0.3\leq\frac{\upsilon_{p}}{\omega_{T}}\leq 0.5.
\end{equation}
In this case, we can calculate from Eqs.~(\ref{Controlled
stationary variances}) and (\ref{xi and xim}) that
$V_{x_M}^{c}\geq V_{p_M}^{c}$, which coincides with the numerical
results in Fig.~\ref{Fig of the squeezing of the normalized
momentum operator}(a) (the green triangular line for
$V_{p_M}^{c}/\hbar$ is below the blue dashed line for
$V_{x_M}^{c}/\hbar$). It means that the fluctuations of the
momentum of the beam are squeezed. {\it More interestingly, the
numerical results in Fig.~\ref{Fig of the squeezing of the
normalized momentum operator}(a) show that the variance of $p_{M}$
is squeezed to be less than the standard quantum limit, i.e.,
$V_{p_M}^{c}<\hbar/2$.} Numerical results in Fig.~\ref{Fig of the
squeezing of the normalized momentum operator}(b) show that the
product $V_{x_M}^{c}V_{p_M}^c$ of the controlled variances is much
smaller than the product $V_{x_M}^{uc}V_{p_M}^{uc}$ of the
uncontrolled variances, which means that the variances of the
position and momentum of the beam could be reduced under the
designed feedback control. Indeed, as shown in Fig.~\ref{Fig of
the squeezing of the normalized momentum operator}(b), the product
of the variances could be reduced to be close to the Heisenberg
uncertainty limit $\hbar^2/4$.

\begin{figure}[h]
\centerline{
\includegraphics[width=3.2in,height=2.5in, clip]{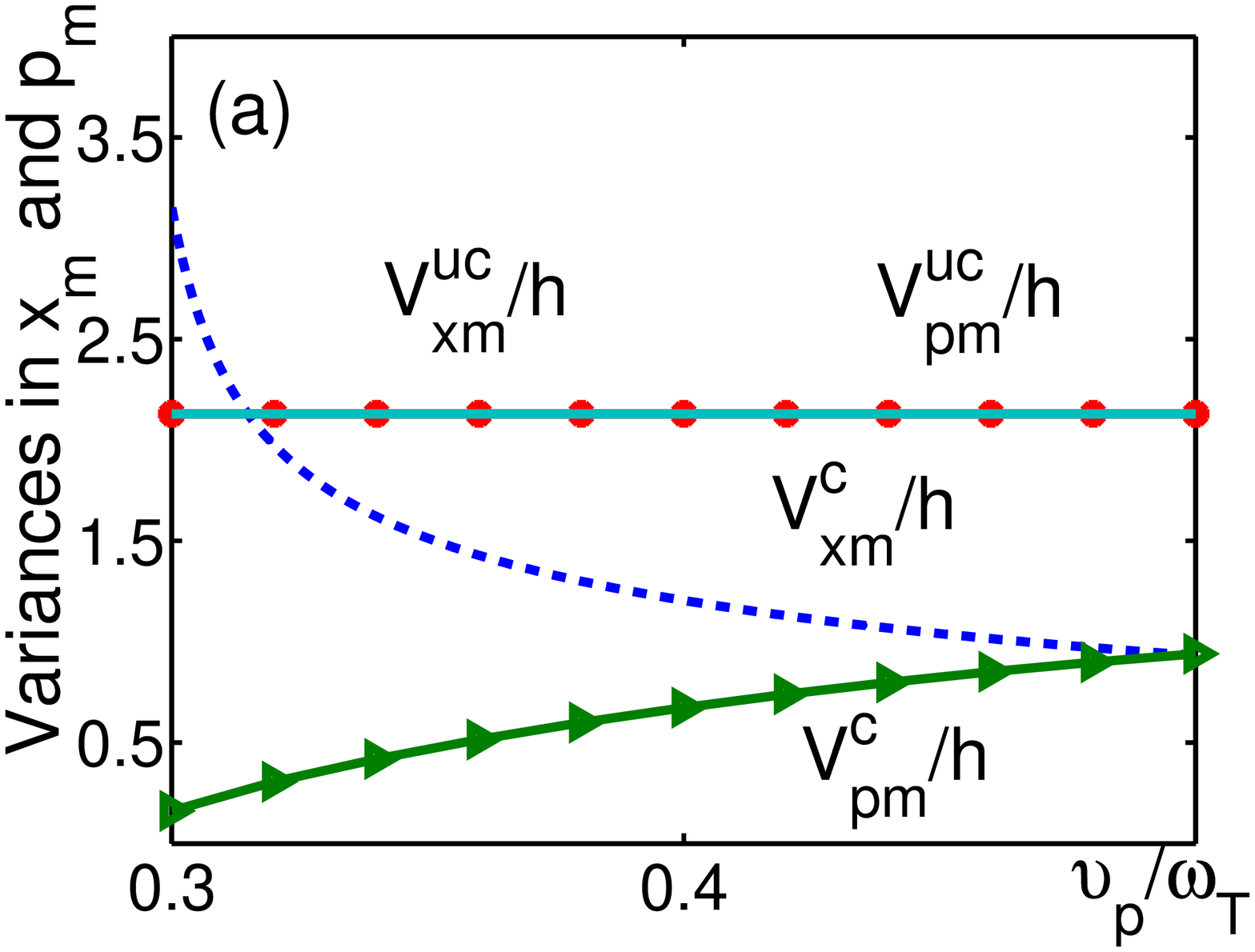}
} \centerline{
\includegraphics[width=3.2in,height=2.5in, clip]{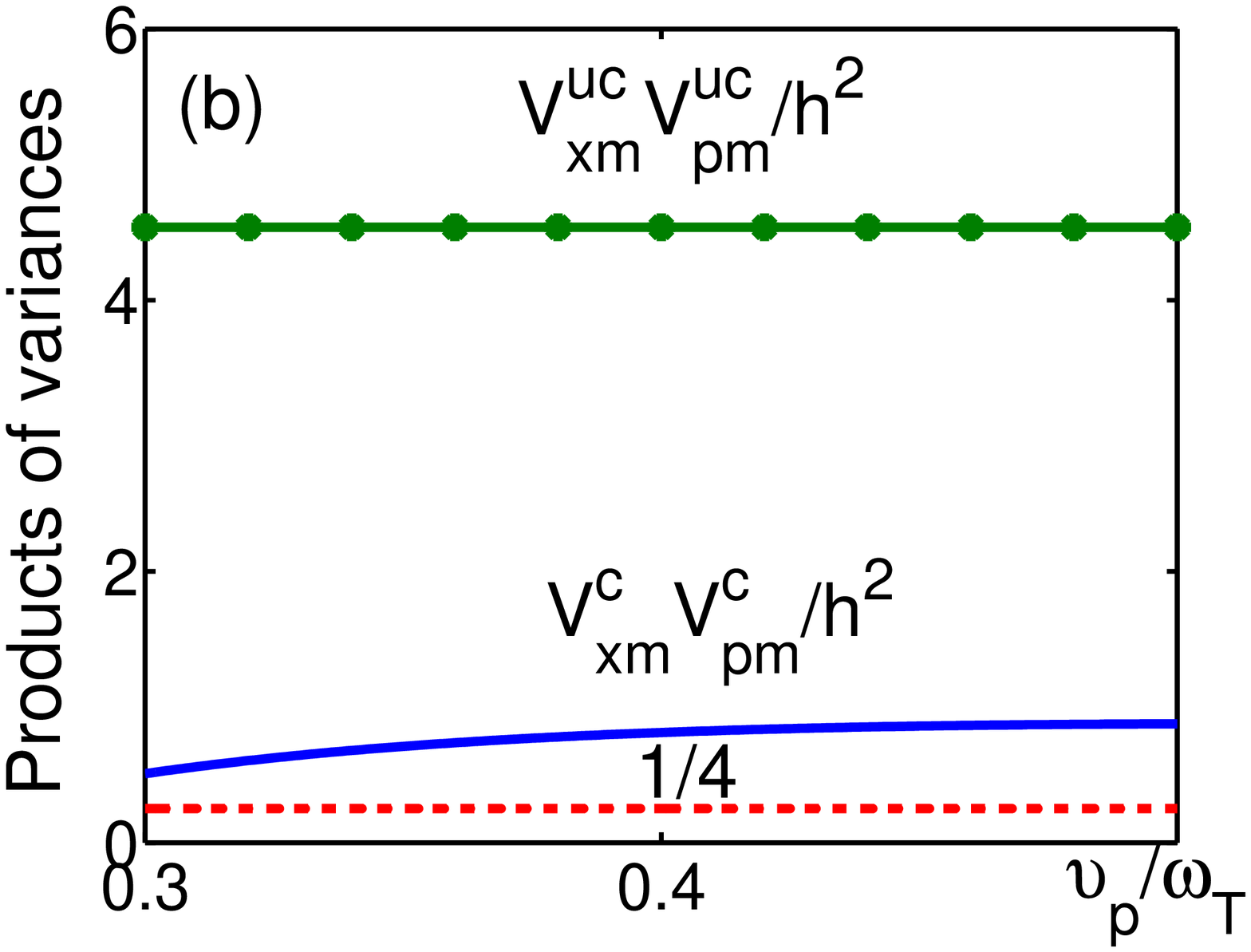}
}\caption{(color online) Squeezing the momentum-fluctuations of
the beam with the parameters given in Eqs.~(\ref{System
parameters1}), (\ref{System parameters2}), and (\ref{Region of the
control parameters for squeezing the momentum}). (a) Variances of
the position and momentum of the beam (in units of $\hbar$) versus
the normalized control parameter
$\tilde{\upsilon}_p=\upsilon_p/\omega_T$. The green line with
triangles and the blue dashed line represent, respectively, the
controlled variances of the position $V_{x_M}^c/\hbar$ and
momentum $V_{p_M}^c/\hbar$ of the beam. The red line with
asterisks and the blue solid line represent the uncontrolled
variances $V_{x_M}^{uc}/\hbar$ and $V_{p_M}^{uc}/\hbar$ of the
beam, which coincide because the beam without control is in a
coherent thermal state with equal variances of position and
momentum. (b) Products of the variances (in units of $\hbar^2$)
versus the normalized control parameter
$\tilde{\upsilon}_p=\upsilon_p/\omega_T$. The green line with
asterisks and the blue solid line denote the uncontrolled
trajectory $V_{x_M}^{uc}V_{p_M}^{uc}/\hbar^2$ and the controlled
trajectory $V_{x_M}^c V_{p_M}^c/\hbar^2$, respectively. The red
dashed line at $1/4$ represents the Heisenberg uncertainty limit
$\hbar^2/4$.}\label{Fig of the squeezing of the normalized
momentum operator}
\end{figure}

\subsection{Cooling}\label{s52}
Further, let us investigate the cooling of the fluctuations of the
nanomechanical beam. The cooling effect can be estimated by the
average photon number of the nanomechanical beam
\begin{eqnarray}\label{Average photon number of the nanomechanical beam}
\bar{n}&=&E_{dW}(\langle b^{\dagger}b\rangle)\nonumber\\
&=&\frac{1}{2\hbar}(V_{x_M}+V_{p_M})-\frac{1}{2}+\frac{1}{2\hbar}(V_{\langle x_M\rangle}+V_{\langle p_M\rangle})\nonumber\\
&&+\frac{1}{2\hbar}(\bar{x}_{M}^2+\bar{p}_{M}^2),
\end{eqnarray}
with
\begin{eqnarray*}
&\bar{x}_{M}=E_{dW}(\langle x_{M}\rangle),\quad\bar{p}_{M}=E_{dW}(\langle p_{M}\rangle),&\\
&V_{\langle x_M\rangle}=E_{dW}(\langle x_{M}\rangle^2)-\bar{x}_{M}^2,&\\
&V_{\langle p_M\rangle}=E_{dW}(\langle
p_{M}\rangle^2)-\bar{p}_{M}^2.&
\end{eqnarray*}
Here, $E_{dW}$ means that the expectations and variances of
$\langle x_{M}\rangle$ and $\langle p_{M}\rangle$ are about the
Wiener noise $dW$.

From the system parameters given in Eqs.~(\ref{System
parameters1}) and (\ref{System parameters2}), it can be verified
that the controlled stationary expectations $\bar{x}_{M}^{c}$,
$\bar{p}_{M}^{c}$ and the classical fluctuations $V^{c}_{\langle
x_M\rangle}$, $V^{c}_{\langle p_M\rangle}$ satisfy
\begin{eqnarray*}
\bar{x}_{M}^{c}=\bar{p}_{M}^{c}=0,\quad V_{\langle
x_M\rangle}^{c}\ll V_{x_M}^c,\quad V_{\langle p_M\rangle}^{c}\ll
V_{p_M}^{c}.
\end{eqnarray*}
Thus, from Eqs.~(\ref{Controlled stationary variances}) and
(\ref{Average photon number of the nanomechanical beam}), the
controlled stationary average photon number can be estimated as:
\begin{eqnarray}\label{The controlled stationary average photon number of the nanomechanical beam}
\bar{n}^{c}&\approx&\frac{1}{2\hbar}(V_{x_M}^{c}+V_{p_M}^{c})-\frac{1}{2}\nonumber\\
&\approx&\frac{1}{4}\left(\sqrt{\frac{\xi}{\eta}}+\frac{1}{\sqrt{\xi\eta}}\right)-\frac{1}{2},
\end{eqnarray}
which, under the parameters given in Eqs.~(\ref{System
parameters1}), (\ref{System parameters2}), (\ref{Region of the
control parameters for squeezing the position}), and (\ref{Region
of the control parameters for squeezing the momentum}), is smaller
than the uncontrolled stationary average photon number
\begin{eqnarray*}
\bar{n}^{uc}\approx\frac{1}{2\hbar}(V_{x_M}^{uc}+V_{p_M}^{uc})-\frac{1}{2}\approx\bar{n}_{M},
\end{eqnarray*}
where $\bar{n}_{M}$ is given in Eq.~(\ref{nbarm}).

Alternatively, we can use an effective temperature $T_{\rm eff}$
to quantify the cooling effect which is defined by:
\begin{eqnarray*}
\bar{n}=\frac{1}{e^{\hbar\omega_{M}/k_{B}T_{\rm eff}}-1},
\end{eqnarray*}
or, equivalently,
\begin{equation}\label{Effective temperature}
T_{\rm eff}=\frac{\hbar\omega_{M}}{k_{B}
\ln\left(\frac{\bar{n}+1}{\bar{n}}\right)}.
\end{equation}
The controlled stationary effective temperature $T_{\rm eff}^c$
can be estimated as follows:
\begin{equation}\label{Effective temperature under control}
T_{\rm
eff}^{c}\approx\frac{\hbar\omega_{M}}{k_{B}\ln\left(\frac{\sqrt{\xi/\eta}+\sqrt{1/\xi\eta}+2}{\sqrt{\xi/\eta}+\sqrt{1/\xi\eta}-2}\right)}.
\end{equation}

We now give some physical interpretations of our cooling strategy.
There are two competing processes that determine the stationary
effective temperature of the nanomechanical beam. The cooling
process is provided by the leakage of the transmission line
resonator, whose energy gap is larger than $k_{B} T$ such that the
thermal excitation from the environment could be negligible.
Energy flows from the beam to the transmission line resonator via
the coupling between them, and then it is dissipated via the
leakage of the transmission line resonator. An opposing heating
process is provided by the thermal excitation of the beam from the
environment. Without applying quantum feedback control on the
transmission line resonator, the cooling process of the beam is
weak compared with the heating process, which leads to the failure
of cooling. When we apply quantum feedback control and adjust the
control parameters to be in the region given by Eq.~(\ref{Region
of the control parameters}), the decay of the beam caused by the
leakage from the transmission line resonator is enhanced to
overwhelm the heating process. Thus, the nanomechanical beam is
effectively cooled.

To show the validity of our proposal, let us show some numerical
examples. The system parameters are chosen as in Eqs.~(\ref{System
parameters1}) and (\ref{System parameters2}), and the feedback
control parameters $\upsilon_{x}$ and $\upsilon_{p}$ are chosen
such that
\begin{equation}\label{Region of the control parameters for cooling}
\upsilon_{x}=0.5\,\omega_T,\quad
0.3\leq\frac{\upsilon_{p}}{\omega_{T}}\leq 1.
\end{equation}
The numerical results in Fig.~\ref{Fig of the cooling of the
nanomechanical beam} show that the average photon number and the
effective temperature of the beam under control are reduced
compared with the uncontrolled case. It means that our strategy
can indeed effectively cool the motion of the beam. With the
parameters given in Eqs.~(\ref{System parameters1}), (\ref{System
parameters2}), and (\ref{Region of the control parameters for
cooling}), the minimum average photon number that can be reached
is about $0.43$ (corresponding to an effective temperature $T_{\rm
eff}^{c}\approx 6.3$ mK). Further calculations show that, if we
increase $\upsilon_{p}/\omega_{T}$ further, the minimum average
photon number that can be reached by our strategy is about $0.35$
(corresponding to an effective temperature $T_{\rm eff}^{c}\approx
5.6$ mK).

\begin{figure}[h]
\centerline{
\includegraphics[width=3.2in,height=2.5in, clip]{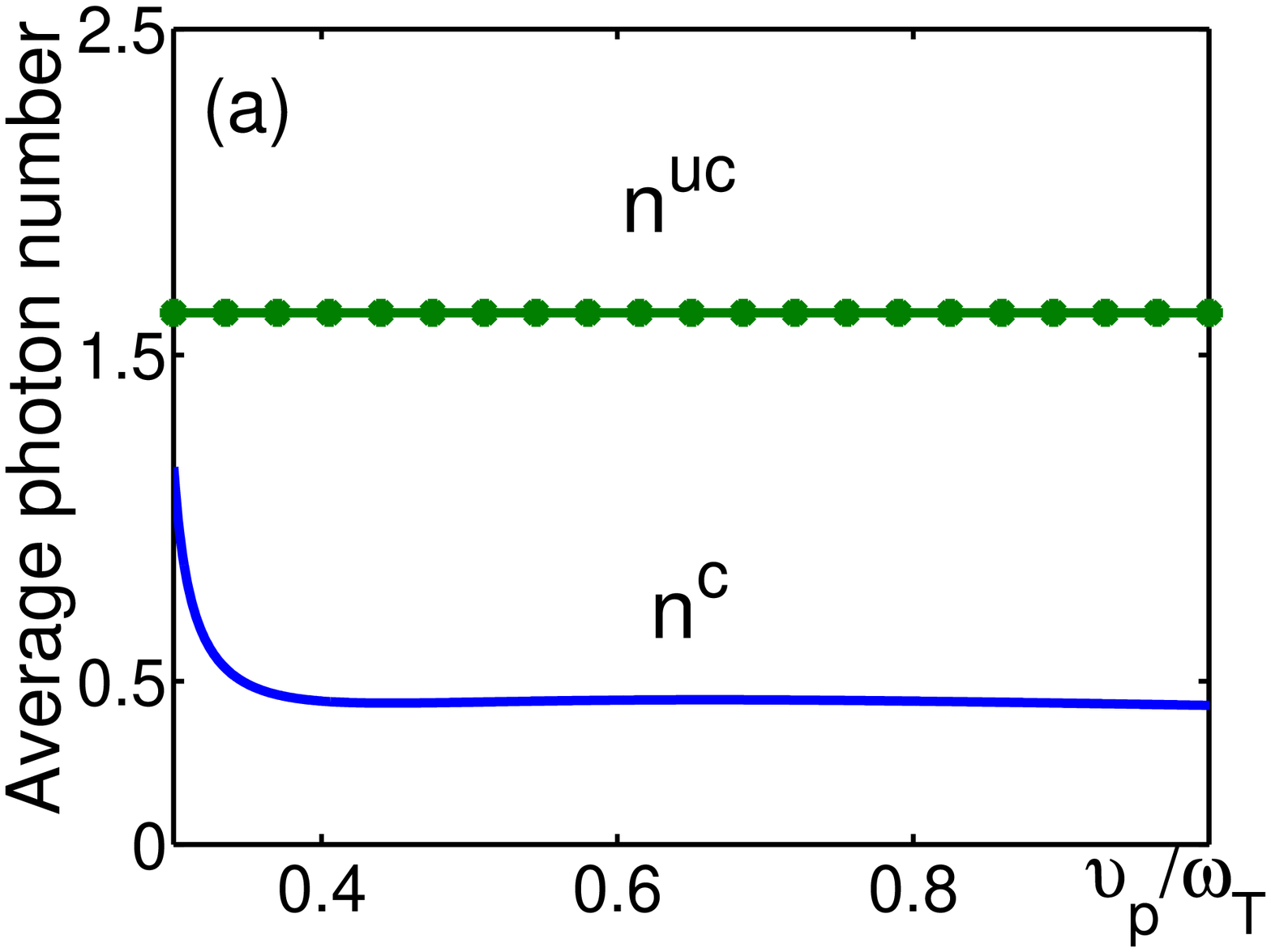}
} \centerline{
\includegraphics[width=3.2in,height=2.5in, clip]{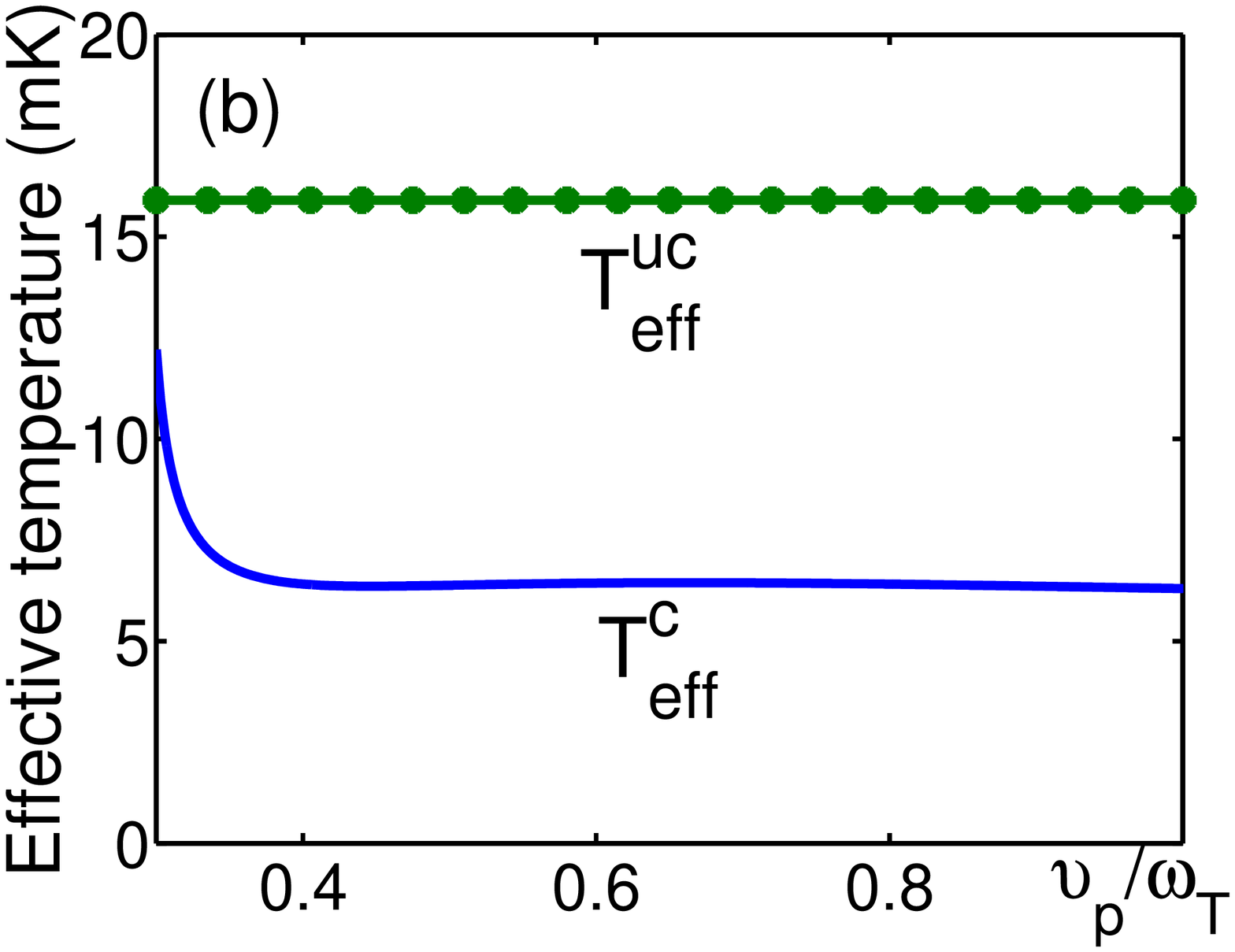}
}\caption{(color online) Cooling the nanomechanical beam for (a)
the average photon number $\bar{n}$, and (b) the effective
temperature $T_{\rm eff}$ versus the normalized control parameter
$\tilde{\upsilon}_p=\upsilon_p/\omega_T$. The blue solid lines
representing the average photon number $n^c$ and effective
temperature $T_{\rm eff}^c$ under control are below the green
lines with asterisks representing the corresponding average photon
number $n^{uc}$ and effective temperature $T_{\rm eff}^{uc}$
without control, which means that the designed feedback control
could effectively cool the fluctuations of the beam induced by
thermal noises.}\label{Fig of the cooling of the nanomechanical
beam}
\end{figure}

\section{Conclusions}\label{s6}

In summary, we have investigated the possibility of using quantum
feedback control to squeeze and cool down the fluctuations of a
nanomechanical beam embedded in a coupled transmission line
resonator--SQUID--mechanical beam quantum circuit. The leakage of
the electromagnetic field from the transmission line resonator is
detected using homodyne measurement, and the measurement output is
then used to design a quantum-feedback-control signal to drive the
electromagnetic field in the transmission line resonator. The
designed quantum-feedback-control protocol indirectly affects the
motion of the beam by the inductive coupling between the
transmission line resonator and the beam via the rf- SQUID. After
adiabatically eliminating the degrees of freedom of the rf-SQUID
and the transmission line resonator, the quantum feedback control
results in a two-photon term in the effective Hamiltonian and
additional damping terms for the beam, which lead to squeezing and
cooling for the beam. By varying the feedback control parameters,
the variance of either the position or momentum of the beam could
be squeezed, and the variance of the momentum of the beam could
even be squeezed to be less than the standard quantum limit
$\hbar/2$. Meanwhile, the average photon number (or, equivalently,
the effective temperature) of the beam could be reduced
effectively by applying control, compared with the uncontrolled
case.

Although the thermal motion of the beam could be effectively
suppressed by the proposed quantum feedback control protocol, our
calculations show that the beam has not achieved the ground state.
Further work will focus on extending our results to explore ways
to further lower the achievable effective temperature or even
attain the ground state of the beam. Another possible direction is
to consider nonlinear effects~\cite{Jacobs2} of the nanomechanical
oscillator, which may affect the achievable cooling temperature
and the squeezing effects induced by the quantum feedback control.
\\[0.2cm]

\begin{center}
\textbf{ACKNOWLEDGMENTS}
\end{center}

We thank Dr. S.~Ashhab for his great help on this work. J. Zhang
would like to thank Dr. N.~Yamamoto for helpful discussions. FN
acknowledges partial support from the National Security Agency
(NSA), Laboratory Physical Science (LPS), Army Research Office
(ARO), National Science Foundation (NSF) grant No. EIA-0130383,
JSPS-RFBR 06-02-91200. J. Zhang was supported by the National
Natural Science Foundation of China under Grant Nos. 60704017,
60635040, 60674039 and the China Postdoctoral Science Foundation.
\\[0.2cm]

\appendix
\section{Derivation of the quantum stochastic differential equation (\ref{Quantum stochastic differential equation of the coupled resonator-SQUID-beam system})}\label{Derivation of the quantum differential equation}
Consider a bosonic model of the environmental degrees of freedom,
and assume that the interactions between the degrees of freedom of
the system and the environmental ones have linear interactions.
Then, under the rotating-wave approximation, the Hamiltonian of
the total system, composed of the degrees of freedom of the system
and the environment, can be expressed as:
\begin{eqnarray*}
H_{\rm tot}&=&H_{\rm eff}+\int\hbar\omega c_{pT}^{\dagger}c_{pT}d\omega+\int\hbar\omega c_{eT}^{\dagger}c_{eT}d\omega\\
&&+\int\hbar\omega c_{eM}^{\dagger}c_{eM}d\omega+\int\hbar\omega c_{eS}^{\dagger}c_{eS}d\omega\\
&&+\hbar\int d\omega\left[g_{pT}^*(\omega)c^{\dagger}_{pT}a+g_{pT}(\omega)a^{\dagger}c_{pT}\right]\\
&&+\hbar\int d\omega\left[g_{eT}^*(\omega)c^{\dagger}_{eT}a+g_{eT}(\omega)a^{\dagger}c_{eT}\right]\\
&&+\hbar\int d\omega\left[g_{eM}^*(\omega)c^{\dagger}_{eM}b+g_{eM}(\omega)b^{\dagger}c_{eM}\right]\\
&&+\hbar M_{\phi}\int d\omega\left[g_{eS}^*(\omega)c^{\dagger}_{eS}\sigma_z+g_{eS}(\omega)\sigma_z c_{eS}\right]\\
&&+\hbar M_{r}\int
d\omega\left[g_{eS}^*(\omega)c^{\dagger}_{eS}\sigma_-+g_{eS}(\omega)\sigma_+
c_{eS}\right],
\end{eqnarray*}
where $c_{pT}$ ($c_{pT}^{\dagger}$), $c_{eT}$
($c_{eT}^{\dagger}$), $c_{eM}$ ($c_{eM}^{\dagger}$), $c_{eS}$
($c_{eS}^{\dagger}$) are, respectively, the annihilation
(creation) operators of different environmental degrees of
freedom, which satisfy
\begin{eqnarray*}
[c_i(\omega),c_j^{\dagger}(\omega')]=\delta_{ij}\delta(\omega-\omega').
\end{eqnarray*}
The subscripts ``pT" and ``eT" represent the environmental degrees
of freedom interacting with the transmission line resonator being
``probed" (pT) and not being probed (eT). The subscripts ``eM" and
``eS" denote the environmental (thus the ``e") degrees of freedom
interacting with the mechanical beam and the SQUID, respectively.

Let $X$ be a system operator, then the Heisenberg equation for $X$
can be written as:
\begin{eqnarray}\label{Differential equation form of the Heisenberg equation of X}
dX&=&-\frac{i}{\hbar}[X,H_{\rm tot}dt]-\frac{1}{2\hbar^2}[[X,H_{\rm tot}dt],H_{\rm tot}dt]\nonumber\\
&=&-\frac{i}{\hbar}[X,H_{\rm eff}]dt+d\,L_{pT}(X)+dL_{eT}(X)\nonumber\\
&&+dL_{eM}(X)+dL_{eS,\,\varphi}(X)+dL_{eS,\,r}(X),\nonumber\\
\end{eqnarray}
where
\begin{eqnarray*}
dL_{\alpha}(X)=-\frac{i}{\hbar}[X,H_{\alpha}dt]-\frac{1}{2\hbar^2}[[X,H_{\alpha}dt],H_{\alpha}dt],
\end{eqnarray*}
and
\begin{eqnarray*}
H_{pT}&=&\hbar\int d\omega\left[g_{pT}^*(\omega)c_{pT}^{\dagger}a+g_{pT}(\omega)a^{\dagger}c_{pT}\right],\\
H_{eT}&=&\hbar\int d\omega\left[g_{eT}^*(\omega)c_{eT}^{\dagger}a+g_{eT}(\omega)a^{\dagger}c_{eT}\right],\\
H_{eM}&=&\hbar\int d\omega\left[g_{eM}^*(\omega)c_{eM}^{\dagger}b+g_{eM}(\omega)b^{\dagger}c_{eM}\right],\\
H_{eS,\varphi}&=&\hbar M_{\phi}\int d\omega\left[g_{eS}^*(\omega)c_{eS}^{\dagger}\sigma_z+g_{eS}(\omega)\sigma_z c_{eS}\right],\\
H_{eS,r}&=&\hbar M_r\int
d\omega\left[g_{eS}^*(\omega)c_{eS}^{\dagger}\sigma_-+g_{eS}(\omega)\sigma_+
c_{eS}\right].
\end{eqnarray*}
Here, we expand $dX$ to second-order differential terms, because
we may meet quantum Wiener increments and the second-order terms
cannot be omitted.

In order to eliminate the environmental degrees of freedom
corresponding to $c_{pT}(\omega)$, we first solve the equation for
$c_{pT}(\omega)$:
\begin{eqnarray*}
\dot{c}_{pT}(\omega)=-\frac{i}{\hbar}[c_{pT}(\omega),H_{\rm
tot}]=-i\omega\,c_{pT}(\omega)-ig_{pT}^*(\omega)a
\end{eqnarray*}
to obtain
\begin{eqnarray}\label{Solution of apT}
c_{pT}(\omega,t)&=&\int_{t_0}^tdt'e^{-i\omega(t-t')}(-ig^*_{pT}(\omega))a(t')\nonumber\\
&&+e^{-i\omega(t-t_0)}c_{pT}(\omega,t_0).
\end{eqnarray}
Further, let us introduce the so-called first Markovian approximation to omit the frequency dependence of the coupling strength~\cite{Buks2}:
\begin{equation}\label{The first Markovian approximation of gpT}
g_{pT}(\omega)=\sqrt{\frac{\gamma_{pT}}{2\pi}}e^{i\phi_{pT}},
\end{equation}
where $\gamma_{pT}$ and $\phi_{pT}$ are independent of $\omega$.

By substituting Eqs. (\ref{Solution of apT}) and (\ref{The first
Markovian approximation of gpT}) into $dL_{pT}(X)$, we have
\begin{eqnarray*}
-\frac{i}{\hbar}[X,H_{pT}dt]&=&\left\{\int d\omega|g_{pT}(\omega)|^2\int_{t_0}^t dt' e^{-i\omega(t-t')}\right.\\
&&\left.(a^{\dagger}(t')[X,a]+a(t')[a^{\dagger},X])\right\}dt\\
&&-ig^*_{pT}\int d\omega e^{i\omega(t-t_0)}c^{\dagger}_{pT}(\omega,t_0)[X,a]dt\\
&&+ig_{pT}[a^{\dagger},X]\int d\omega e^{-i\omega(t-t_0)}c_{pT}(\omega,t_0)dt\\
&=&\frac{\gamma_{pT}}{2}\left(a^{\dagger}[X,a]-a[X,a^{\dagger}]\right)dt\\
&&+\left(-i\sqrt{\gamma_{pT}}e^{-i\phi_{pT}}d\tilde{A}_{\rm in}^{\dagger}\right)[X,a]\\
&&+[a^{\dagger},X]\left(i\sqrt{\gamma_{pT}}e^{i\phi_{pT}}d\tilde{A}_{\rm
in}\right),
\end{eqnarray*}
where
\begin{eqnarray*}
d\tilde{A}_{\rm in}=\left(\frac{1}{\sqrt{2\pi}}\int d\omega
e^{-i\omega(t-t_0)}c_{pT}(\omega,t_0)\right)dt
\end{eqnarray*}
is the input quantum noise such that
\begin{eqnarray*}
&d\tilde{A}_{\rm in}d\tilde{A}_{\rm in}^{\dagger}=(\bar{n}(\omega_{T})+1)dt,&\\
&d\tilde{A}_{\rm in}^{\dagger}d\tilde{A}_{\rm in}=\bar{n}(\omega_{T})dt,&\\
&d\tilde{A}_{\rm in}d\tilde{A}_{\rm in}=d\tilde{A}_{\rm
in}^{\dagger}d\tilde{A}_{\rm in}^{\dagger}=0,&
\end{eqnarray*}
and
\begin{eqnarray*}
\bar{n}(\omega)=\frac{1}{e^{\hbar\omega/k_{B} T}-1}.
\end{eqnarray*}
Further, we have
\begin{eqnarray*}
-\frac{1}{2\hbar^2}[[X,H_{pT}dt],H_{pT}dt]&=&-\frac{\gamma_{pT}}{2}\bar{n}(\omega_{T})[[X,a],a^{\dagger}]dt\\
&&-\frac{\gamma_{pT}}{2}(\bar{n}(\omega_{T})+1)[[X,a^{\dagger}],a]dt.
\end{eqnarray*}
From the above analysis, it can be calculated that
\begin{eqnarray*}
d\,L_{pT}(X)&=&-\frac{i}{\hbar}[X,H_{pT}dt]-\frac{1}{2\hbar^2}[[X,H_{pT}],H_{pT}dt]\\
&=&\frac{\gamma_{pT}}{2}(\bar{n}(\omega_{T})+1)\left(a^{\dagger}[X,a]+[a^{\dagger},X]a\right)\\
&&+\frac{\gamma_{pT}}{2}\bar{n}(\omega_{T})\left(a[X,a^{\dagger}]+[a,X]a^{\dagger}\right)\\
&&+\sqrt{\gamma_{pT}}dA_{\rm
in}^{\dagger}[X,a]+\sqrt{\gamma_{pT}}[a^{\dagger},X]dA_{\rm in},
\end{eqnarray*}
where $dA_{\rm in}=ie^{i\phi_{pT}}d\tilde{A}_{\rm in}$.

With the same analysis, we can calculate $dL_{eT}(X)$,
$dL_{eM}(X)$, $dL_{eS,\varphi}(X)$, and $dL_{eS,r}(X)$.
Furthermore, under the condition that
$\hbar\omega_{S},\hbar\omega_{T}\gg k_{B} T$, we have
$\bar{n}(\omega_{S}),\,\bar{n}(\omega_{T})\approx 0$. Thus, by
substituting the above results into Eq.~(\ref{Differential
equation form of the Heisenberg equation of X}), averaging over
the fluctuations caused by the thermal noises, and assuming that
\begin{eqnarray*}
\gamma_{T}=\gamma_{pT}+\gamma_{eT},\quad\eta=\frac{\gamma_{pT}}{\gamma_{pT}+\gamma_{eT}},
\end{eqnarray*}
we can obtain the quantum stochastic differential equation
(\ref{Quantum stochastic differential equation of the coupled
resonator-SQUID-beam system}).

In order to calculate the measurement output of the homodyne
detection, let us recall that the input and output detection
noises should be
\begin{eqnarray*}
d\tilde{A}_{\rm in}&=&\left(\frac{1}{\sqrt{2\pi}}\int d\omega e^{-i\omega(t-t_0)}c_{pT}(\omega,t_0)\right)dt,\\
d\tilde{A}_{\rm out}&=&\left(\frac{1}{\sqrt{2\pi}}\int d\omega
e^{-i\omega(t-t_0)}c_{pT}(\omega,t_1)\right)dt,
\end{eqnarray*}
where the time $t_0$ is an instant before the measurement
commences, and the time $t_1$ is another instant after the
measurement has finished. The measurement output is related to
$d\tilde{A}_{\rm out}$ by:
\begin{eqnarray*}
dY_{t}=e^{-i\phi_{LO}}d\tilde{A}_{\rm
out}^{\dagger}+e^{i\phi_{LO}}d\tilde{A}_{\rm out},
\end{eqnarray*}
where $\phi_{LO}$ is an adjustable phase introduced by the local
oscillator of the homodyne detection. From Eq.~(\ref{Solution of
apT}), we have
\begin{eqnarray*}
c_{pT}(\omega,t)&=&-i\sqrt{\gamma_{pT}}e^{-i\phi_{pT}}\frac{1}{\sqrt{2\pi}}\int_{t_0}^tdt'e^{-i\omega(t-t')}a(t')\nonumber\\
&&+e^{-i\omega(t-t_0)}c_{pT}(\omega,t_0)\\
&=&-i\sqrt{\gamma_{pT}}e^{-i\phi_{pT}}\frac{1}{\sqrt{2\pi}}\int_{t_1}^tdt'e^{-i\omega(t-t')}a(t')\nonumber\\
&&+e^{-i\omega(t-t_1)}c_{pT}(\omega,t_1).
\end{eqnarray*}
Thus, it can be calculated that
\begin{eqnarray*}
d\tilde{A}_{\rm out}-d\tilde{A}_{\rm
in}=-i\sqrt{\gamma_{pT}}e^{-i\phi_{pT}}a(t)dt,
\end{eqnarray*}
from which it can be shown that
\begin{eqnarray*}
dY_{t}&=&i\sqrt{\gamma_{pT}}a^{\dagger}e^{i(\phi_{pT}-\phi_{LO})}-i\sqrt{\gamma_{pT}}ae^{-i(\phi_{pT}-\phi_{LO})}\\
&&+e^{-i\phi_{LO}}d\tilde{A}_{\rm
in}^{\dagger}+e^{i\phi_{LO}}d\tilde{A}_{\rm in}.
\end{eqnarray*}
By setting $\phi_{LO}=\phi_{pT}+\pi/2$, we have
\begin{eqnarray*}
dY_{t}&=&\sqrt{\gamma_{pT}}(a^{\dagger}+a)+\left(i e^{i\phi_{pT}}d\tilde{A}_{\rm in}-ie^{-i\phi_{pT}}d\tilde{A}_{\rm in}^{\dagger}\right)\\
&=&\sqrt{\eta\gamma_{T}}(a^{\dagger}+a)+(dA_{\rm in}+dA_{\rm
in}^{\dagger}).
\end{eqnarray*}

\section{Derivation of the reduced stochastic master equation (\ref{Stochastic master equation of the beam})}\label{Derivation of the reduced stochastic master equation for the beam}

Under the semiclassical approximation, we can obtain
Maxwell-Bloch-type equations from the stochastic master equation
(\ref{Stochastic master equation for the beam and the transmission
line resonator}) for the coupled beam-SQUID-resonator system (see,
e.g., Ref.~\cite{Mabuchi2}). Further, in the large-detuning regime
(see Eq.~(\ref{Large detuning assumption})), we have
\begin{equation}\label{Small frequency shift condition}
\frac{g_{MS}^2}{\Delta_{MS}},\,\frac{g_{ST}^2}{\Delta_{ST}},\,g_{MT}\ll\omega_M,\,\omega_T,\,\omega_S,
\end{equation}
where
\begin{equation}\label{Reduced coupling strength between the beam and the resonator}
g_{MT}=g_{MS}g_{ST}\left(\frac{1}{\Delta_{MS}}+\frac{1}{\Delta_{ST}}\right).
\end{equation}
Then, we can omit the frequency shifts of the beam, the rf-SQUID,
and the transmission line resonator induced by the coupling
between them. Under this condition, the Maxwell-Bloch-type
equations for the coupled system obtained from the stochastic
master equation (\ref{Stochastic master equation for the beam and
the transmission line resonator}) can be expressed as:
\begin{eqnarray}\label{Maxwell-Bloch-type equation for the coupled beam-SQUID-resonator system}
\dot{\langle\sigma_x\rangle}&=&-\omega_S\langle\sigma_y\rangle-2\gamma_S\left(M_{\phi}^2+\frac{1}{4}M_r^2\right)\langle\sigma_x\rangle,\nonumber\\
\dot{\langle\sigma_y\rangle}&=&\omega_S\langle\sigma_x\rangle-2\gamma_S\left(M_{\phi}^2+\frac{1}{4}M_r^2\right)\langle\sigma_y\rangle,\nonumber\\
\dot{\langle\sigma_z\rangle}&=&-\gamma_SM_r^2\langle\sigma_z\rangle-\gamma_SM_r^2,\nonumber\\
\dot{\langle b\rangle}&=&-i\omega_M\langle
b\rangle+g_{MT}\langle\sigma_z\rangle\langle
a\rangle-\frac{\gamma_{M}}{2}\langle b\rangle,\nonumber\\
d\langle a\rangle&=&-i\omega_T\langle a\rangle
dt-g_{MT}\langle\sigma_z\rangle\langle b\rangle
dt-\frac{\gamma_T}{2}\langle a\rangle dt-i u(t)dt\nonumber\\
&&+\sqrt{\frac{\eta\gamma_T}{\hbar}}\left(V_{x_T}+i
C_{x_Tp_T}-\frac{\hbar}{2}\right)dW,
\end{eqnarray}
where $dW$ has been given in Eq.~(\ref{Wiener increment});
\begin{eqnarray*}
V_{x_T}=\langle x_T^2\rangle-\langle x_T\rangle^2,\quad
V_{p_T}=\langle p_T^2\rangle-\langle p_T\rangle^2
\end{eqnarray*}
are the variances of the normalized position and momentum
operators of the transmission line resonator given by
Eq.~(\ref{Normalized position and momentum of the transmission
line resonator}); and
\begin{eqnarray*}
C_{x_T
p_T}=\left\langle\frac{x_Tp_T+p_Tx_T}{2}\right\rangle-\langle
x_T\rangle\langle p_T\rangle
\end{eqnarray*}
is the corresponding symmetric covariance. Under the semiclassical
approximation and the condition (\ref{Small frequency shift
condition}), $V_{x_T}$, $V_{p_T}$ and $C_{x_Tp_T}$ can be given by
the following equations:
\begin{eqnarray}\label{Equations for the variances and covariance of the transmission line resonator}
\dot{V}_{x_T}&=&-\gamma_T V_{x_T}+2\omega_T C_{x_T
p_T}+\frac{\hbar\gamma_T}{2}\nonumber\\
&&-2\eta\gamma_T\left(V_{x_T}-\frac{\hbar}{2}\right)^2,\nonumber\\
\dot{V}_{p_T}&=&-\gamma_T V_{p_T}-2\omega_T
C_{x_Tp_T}+\frac{\hbar\gamma_T}{2}\nonumber\\
&&-2\eta\gamma_T
C_{x_Tp_T}^2,\nonumber\\
\dot{C}_{x_Tp_T}&=&-\gamma_TC_{x_Tp_T}+\omega_T V_{p_T}-\omega_T
V_{x_T}\nonumber\\
&&-2\eta\gamma_T\left(V_{x_T}-\frac{\hbar}{2}\right)C_{x_Tp_T}.
\end{eqnarray}

By substituting the feedback control~(\ref{Linear feedback
control}) into Eq.~(\ref{Maxwell-Bloch-type equation for the
coupled beam-SQUID-resonator system}), we can replace the last
equation in (\ref{Maxwell-Bloch-type equation for the coupled
beam-SQUID-resonator system}) by the following equation:
\begin{eqnarray}\label{Equation of the expectation of aT after substituting the feedback control}
d\langle a\rangle&=&-i\omega_{T}\langle a\rangle dt-g_{MT}\langle\sigma_z\rangle\langle b\rangle dt-\frac{\gamma_{T}}{2}\langle a\rangle dt\nonumber\\
&&+i\upsilon_{x}\langle a+a^{\dagger}\rangle dt-i\upsilon_{p}\langle -ia+ia^{\dagger}\rangle dt\nonumber\\
&&+\sqrt{\frac{\eta\gamma_{T}}{\hbar}}\left(V_{x_T}+iC_{x_T
p_T}-\frac{\hbar}{2}\right)dW.
\end{eqnarray}

If the damping rates $\gamma_S$ and $\gamma_{T}$ of the rf-SQUID
and the transmission line resonator are large enough such that
\begin{equation}\label{Adiabatic elimination condition}
\gamma_{S},\,\gamma_{T}\gg\gamma_{M}\bar{n}_{M},
\end{equation}
we can adiabatically eliminate~\cite{Walls,Steck} the degrees of
freedom of the rf-SQUID and the transmission line resonator to
obtain the reduced equation of the beam. In fact, in this case, we
can obtain the following stationary variances from
Eq.~(\ref{Equations for the variances and covariance of the
transmission line resonator}):
\begin{eqnarray*}
V_{x_T}=V_{p_T}=\frac{\hbar}{2},\,\,\, C_{x_T p_T}=0,
\end{eqnarray*}
from which it can be verified that the fluctuation in
Eq.~(\ref{Equation of the expectation of aT after substituting the
feedback control}) will tend to zero. Then, in the Heisenberg
picture, one finds from the stationary solution of
Eqs.~(\ref{Maxwell-Bloch-type equation for the coupled
beam-SQUID-resonator system}) and (\ref{Equation of the
expectation of aT after substituting the feedback control}) that
\begin{equation}\label{Relationship between aT and am}
\sigma_z\sim-1,\quad a\sim C_1 b+ C_2 b^{\dagger},
\end{equation}
where $C_1$, $C_2$ are given by Eq.~(\ref{C1C2xi}). Substituting
Eq.~(\ref{Relationship between aT and am}) into
Eq.~(\ref{Stochastic master equation for the beam and the
transmission line resonator}), we can obtain the reduced
stochastic master equation~(\ref{Stochastic master equation of the
beam}) for the nanomechanical beam.
\\[0.2cm]

\end{document}